\documentclass[preprint, 11pt]{elsarticle}
\usepackage[utf8]{inputenc}
\usepackage[english]{babel}
\usepackage{textcomp}

\usepackage{bm}

\usepackage{amsthm, mathtools, amsmath, amssymb, array}
\usepackage{subcaption}
\usepackage{siunitx}

\usepackage{tabularx}
\usepackage{diagbox}
\usepackage{url}
\usepackage{hyperref}
\usepackage[capitalise]{cleveref}
\crefrangelabelformat{equation}{(#3#1#4-#5#2#6)}
\usepackage[dvipsnames]{xcolor}
\usepackage{graphicx}
\usepackage{placeins, afterpage}
\usepackage{setspace}

\journal{Acta Astronautica}

\begin{document}
	

\begin{frontmatter}
\title{Target selection for Near-Earth Asteroids in-orbit sample collection missions}
\author{Mirko Trisolini\corref{cor1}}
\ead{mirko.trisolini@polimi.it}
\address{Politecnico di Milano, Via La Masa 34, 20156, Milano, Italy}
\author{Camilla Colombo}
\ead{camilla.colombo@polimi.it}
\address{Politecnico di Milano, Via La Masa 34, 20156, Milano, Italy}
\author{Yuichi Tsuda}
\ead{tsuda.yuichi@jaxa.jp}
\address{Institute of Space and Astronautical Science (ISAS)/Japan Aerospace Exploration Agency (JAXA), 3-1-1 Yoshinodai, Chuo-ku, Sagamihara, Kanagawa 252-5210, Japan}
\cortext[cor1]{Corresponding author}

\begin{abstract}
This work presents a mission concepts for in-orbit particle collection for sampling and exploration missions towards Near-Earth asteroids. Ejecta is generated via a small kinetic impactor and two possible collection strategies are investigated: collecting the particle along the anti-solar direction, exploiting the dynamical features of the L$_2$ Lagrangian point or collecting them while the spacecraft orbits the asteroid and before they re-impact onto the asteroid surface. Combining the dynamics of the particles in the Circular Restricted Three-Body Problem perturbed by Solar Radiation Pressure with models for the ejecta generation, we identify possible target asteroids as a function of their physical properties, by evaluating the potential for particle collection.
\end{abstract}


\end{frontmatter}

\section{Introduction}
\label{sec:intro}
Space exploration missions to asteroids have always drawn the attention of the scientific and engineering community given the challenges they pose and the possibility they present to further our knowledge of the Solar System. Asteroids carry fundamental information on the evolution of our Solar System. They are rich in valuable resources such as metals, silicates, and water, which could be exploited through future asteroid mining missions \cite{hein2020techno,xie2021target}, and enable long-duration mission self-sustaining. The physical composition of asteroids is varied and, in most cases, poorly understood; it can be significantly improved collecting and studying their samples. Improving our knowledge, we can better target asteroids to be exploited for mining and increase the efficiency of asteroid deflection missions. Several missions have visited asteroids and other small bodies; however, only few have orbited, landed, or impacted on them. Examples are JAXA's missions Hayabusa \citep{kawaguchi2008hayabusa} and Hayabusa2, which has recently entered its extended phase during which it will perform a flyby wit asteroid 2001 CC-21 and rendezvous with the fast rotating asteroid 1998 KY26 \citep{tsuda2013system,tsuda2019hayabusa2,tsuda2020hayabusa2}. ESA's Rosetta mission visited comet 67P/Churyumov–Gerasimenko \citep{witte2016rosetta,ulamec2016rosetta}, and NASA's Deep Impact comet Temple1 \citep{blume2003deep}. More recently, NASA's OSIRIS-REx recovered samples from asteroid Bennu \citep{lauretta2017osiris} and has been extended as mission OSIRIS-APEx, which will visit asteroid Apophis \citep{dellagiustina2022osiris}. In November 2021 the DART mission \citep{2019adamsDART} was launched and will impact the Dimorphos, the secondary asteroid of the Didymos binary system, to test technologies for future asteroid deflection missions. ESA's HERA mission \citep{dellagiustina2022osiris} will follow DART to inspect the aftermath of the impact on Dimorphos with the objective to validate the technology for asteroid deflection.

This work is part of a larger project that studies the possibility of performing sample collection directly in orbit \citep{latino2019iac,latino2019thesis,trisolini2021ejecta,trisolini2022scitech}, as an alternative strategy to landing or touch down. In fact, landing and touchdown are among the most challenging aspects of sample-collection missions \citep{ulamec2014landing}. In addition, there may be circumstances where landing or touch down operations may be too challenging, for example due to conditions of the terrain, or complex, for example for asteroids with high rotational speeds. Starting from the heritage of JAXA's missions Hayabusa and Hayabusa2, the project focuses on a mission concepts in which we generate fragments around an asteroid via a small kinetic impactor, similarly to Hayabusa2, and we then try to collect them while orbiting. Such a collection mechanism relies on the knowledge of the dynamical behaviour of the small generated fragments orbiting the asteroid, which is influenced by the interaction between the ejecta generation process and the dynamical environment of the asteroid. As part of this project, this work presents a preliminary analysis devoted at identifying possible target asteroids as a function of their characteristics. The evaluation of the target asteroids relies on the study of the evolution of ejecta particles in the context of the circular restricted three body problem (CR3BP) including SRP. The ejection of the particles from the surface of the asteroid is assumed to be obtained via a small kinetic impactor and the relevant distributions of the particles’ size and ejection conditions obtained via an ejecta distribution model derived by experimental correlations \citep{sachse2015correlation,housen2011ejecta,holsapple2007crater}.

Several previous works have studied the dynamical behaviour of small particles around asteroids and comets. In \cite{scheeres2002fate}, Scheeres discusses the fate of ejecta, identifying five classes of possible dynamical evolutions: immediate re-impact, eventual re-impact after at leas one periapsis passage, stable motion, eventual escape after at least one periapsis passage, and immediate escape. These subdivision of ejecta fates is of interest in this project as a some of these types of dynamical behaviour or their subsets can be more interesting for in-orbit particle collection. In \cite{scheeres2000temporary}, Scheeres and Marzari analyse the conditions that lead to quasi-stable orbits after an impact event applied to the specific case of the mission Deep Impact on comet Tempel1 \citep{blume2003deep}. In their work, they apply an analytical methodology developed by Richter \citep{richter1995stability} that considers only the central gravity field and the SRP contribution. Higher order gravity terms and tidal forces are neglected and they are considered to have a smaller influence in the statistical representation of the ejecta fate. This work shows that only certain combinations of particle diameter and ejection speed can lead to quasi-stable orbit, and also that the contribution of the ejection direction can be highly influential. Therefore, in this work, we include the contribution of both the ejection location and ejection direction. Other recent works have focused on the numerical propagation of the ejecta after an impact event. Particularly, in \citep{yu2017ejecta,yu2018ejecta,rossi2022dynamical} the impact of the DART spacecraft on the secondary of a binary system is studied. In this analyses, high fidelity model are used for the propagation of the ejecta trajectories and their global fate evaluated studying a representative samples of particles. Similar high fidelity analyses have been used to verify the risk for impact on Hayabusa2's mission to Ryugu \citep{soldini2017assessing}. Other work, instead have studied the possibility of the impact ejecta to be trapped into periodic orbits under the influence of Solar Radiation Pressure and higher order gravity terms \citep{pinto2020}.
Most of these analyses are dedicated to understanding the fate of an impact event for a specific mission, with very constrained impact characteristics, or to the dynamical analysis about a specific small body. In this work, we are studying a new mission concept that, at the current stage of development, does not have a dedicated target or a specific mission scenario. As a consequence, this work is an effort to define preliminary concepts for in-orbit collection strategies and to identify potentially meaningful targets by using simplified analyses on a large number of possible candidate asteroids. Once potential targets are identified, higher fidelity models can be used to refine the analysis for in-orbit particle collection.

\bigbreak
The objective of this work is therefore to identify potential targets for in-orbit particle collection missions to asteroids. In this context, we introduce two possible scenarios for in-orbit particle collection. A first scenario that leverages the dynamical peculiarities of the Lagrangian point L$_2$, where particles of given size and speed will tend to pass and produce favourable conditions for the collection \citep{latino2019iac,latino2019thesis}. A second scenario that considers the availability of particles in the neighbourhood of the asteroid; these particles can either be on stable or re-impacting orbits so that they can remain available for collection long enough for the spacecraft to collect them. For both these scenarios, we study what targets properties, as size, density, material type and material strength, are the most promising for an effective collection mission and what are the properties that, instead, lead to infeasible mission scenarios. To do so, a Figure of Merit (FOM) is proposed, which represents the number of potential particles available for collection for both the scenarios as a function of the target asteroid properties (i.e., size, density, material type and strength). The key aspect of this work is the combination between the dynamical evolution of the particles and the modelling of the impact ejecta in order to estimate the number of potentially collectable particles. In an effort to have a more general understanding of the in-orbit collection scenarios, we compare different target asteroids based on their size and density, which can be derived and inferred from ground observations \citep{hasegawa2018physical}. Additionally we compare three different types of materials and several strength levels to understand the sensitivity of the collection scenarios to these parameters that have and intrinsically higher level of uncertainty. Finally, we perform a preliminary analysis of the risk the ejected particles pose to the spacecraft by estimating the number of potentially hazardous impacts.

\cref{sec:dynamics} describes the dynamical environment used throughout the work, \cref{sec:ejecta_model} contains the description of the adopted ejecta model, and \cref{sec:collection_strategies} defines the collection strategies and the target evaluation procedures. Finally, \cref{sec:results} shows the evaluation of the target as function of the asteroid characteristics and the identification of a shortlist of possible candidates.

\section{Dynamical model}
\label{sec:dynamics}
The dynamical model used in this work is the Circular Restricted Three-Body Problem (CR3BP) perturbed by Solar Radiation Pressure (SRP). In this model, the gravitational effect of the third body (the ejected particles in our case) is considered negligible with respect to the two primary bodies (the Sun and an asteroid in our case) that move on a Keplerian circular orbit. A rotating (synodic) reference frame is adopted, which follows the motion of the asteroid in its circular orbit around the Sun. Such reference frame is centred in the centre of mass of the asteroid, its x-axis points in the anti-solar direction at all times, its z-axis follows the direction of the asteroid's orbit angular momentum, and the y-axis completes the right-hand orthogonal frame. The CR3BP is traditionally expressed in non-dimensional units by setting the unit of mass equal to the total mass of the system and the unit of length to be the semi-major axis of the orbit of the two primaries \citep{scheeres2016orbital,vallado2001fundamentals}. Following \citep{latino2019thesis}, the equations of motion of the CR3BP perturbed by SRP in a synodic frame take the form:

\begin{equation}  \label{eq:eom}
	\begin{cases}
		\ddot{x} - 2\dot{y} = x + 1 - \mu - \frac{(1 - \beta)(1 - \mu)(x + 1)}{r^3_{\rm sp}} - \frac{\mu}{r^3_{\rm ap}} x \\
		\ddot{y} + 2\dot{x} = - \left[ \frac{(1 - \beta)(1 - \mu)}{r^3_{\rm sp}} + \frac{\mu}{r^3_{\rm ap}} - 1 \right] y \\
		\ddot{z} = - \left[ \frac{(1 - \beta)(1 - \mu)}{r^3_{\rm sp}} + \frac{\mu}{r^3_{\rm ap}} \right] y
	\end{cases}
\end{equation}

where $x$, $y$, and $z$ are the non-dimensional particle positions with respect to the centre of the asteroid in the rotating frame, while $r_{\rm sp}$ and $r_{\rm ap}$ are the distances between the Sun and the particle and the asteroid and the particle, respectively:

\begin{align}
	r_{\rm sp} &= \sqrt{\left( x + 1 \right)^2 + y^2 + z^2} \\
	r_{\rm ap} &= \sqrt{x^2 + y^2 + z^2}
\end{align}

The mass parameter, $\mu$, represents the normalised asteroid mass and has the following expression:

\begin{equation}  \label{eq:mass_parameter}
	\mu = \frac{\mu_A}{\mu_A + \mu_S}
\end{equation}

where $\mu_A$ and $\mu_S$ are the gravitational parameter of the asteroid and the Sun, respectively. The lightness parameter, $\beta$, represents the ratio between the SRP acceleration and the gravitation of the Sun, and has the following expression \citep{latino2019thesis,latino2019iac}:

\begin{equation}  \label{eq:lightness_param}
	\beta = \frac{3}{2} \frac{P_0}{c} \frac{AU^2}{\mu_S} \frac{c_R}{\rho_p d_p}
\end{equation}

where $c_R$ is the reflectivity coefficient (0 for translucent particles, 1 for black bodies, and 2 for reflective particles), $P_0$ = 1367 W/m$^2$ is the solar flux at 1 AU, $c$ is the speed of light, AU is the astronomical unit, $\rho_p$ is the particle density, and $d_p$ the particle diameter. This expression is representative of the so called "cannon ball" model, in which the effect of the SRP acceleration is assumed to act uniformly on an equivalent cross-section, $S$, and always directed along the Sun-particle line \citep{scheeres2016orbital}. In this work, we consider the motion of ejecta. The ejecta particles are modelled as spheres ($S=\pi d_p^2 / 4$) of constant density and possess the same characteristics of the parent asteroid, i.e., same density and reflectivity coefficient. \cref{tab:beta} shows relevant values of the lightness parameters for a combination of target density and particle size of interest for this work.

\begin{table}[htb!]
	\centering
	\caption{\label{tab:beta} Ranges of lightness parameters, $\beta$, as a function of the target density and particle diameter. The reflectivity coefficient is $c_R =$ 0.13, which is the average albedo of NEAs in the NASA Small Body Database \citep{nasa_sbdb}. }
	\begin{tabular}{l|ccc}
		\hline
		\diagbox{$d_p$}{$\rho_p$} & 1.0 g/cm$^3$ & 2.6 g/cm$^3$ & 5.0 g/cm$^3$ \\
		\hline
		0.1 mm		& \num{2.3e-2} 	& \num{8.87e-3}	& \num{4.6e-3} \\
		1 mm		& \num{2.3e-3} 	& \num{8.87e-4}	& \num{4.6e-4} \\
		2 mm		& \num{1.15e-3} 	& \num{4.44e-4} 	& \num{2.3e-4} \\
		\hline
	\end{tabular}
\end{table}

The system of \cref{eq:eom} admits an integral of motion, $C$, known as the Jacobi constant:

\begin{equation}  
	\begin{aligned}  \label{eq:jacobi}
		C = 2U - V^2 &= \left( x^2 + y^2 \right) + 2 \left(1 - \mu \right) x + \frac{2(1-\beta)(1-\mu)}{r_{\rm sp}} + \frac{2\mu}{r_{\rm ap}} + \\
		&+ \left(1 - \mu \right)^2 - \left( \dot{x}^2 + \dot{y}^2 + \dot{z}^2 \right)
	\end{aligned}
\end{equation}

where $U$ is the potential energy of the system and $V = \sqrt{\dot{x}^2 + \dot{y}^2 + \dot{z}^2}$ is the kinetic energy \citep{scheeres2016orbital}. Given that $V \geq 0$, from \cref{eq:jacobi} follows the inequality $2U - C \geq 0$, which can be used to define the regions of accessible and forbidden motion of the particles. The boundaries of these regions are represented by the Zero-velocity curves (ZVC) \citep{szebehely2012theory}, which bound the motion of a particle with a given energy, $C$. It can be shown that the system of \cref{eq:eom} has five equilibrium points \citep{vallado2001fundamentals,szebehely2012theory}, also called Libration points or Lagrangian points. In this work, we are mainly interested in the L$_1$ and L$_2$ collinear libration points, which are located along the line connecting the Sun and the asteroid. Specifically, the equilibrium point L$_1$ is located between the Sun and the asteroid, while the equilibrium point L$_2$ after the asteroid, in the anti-solar direction (on the positive side of x-axis of the synodic frame). The location of the equilibrium points is influenced by the strength of the SRP acceleration. An increase in SRP causes the L$_1$ to move closer to the Sun, while the L$_2$ moves closer to the asteroid. In this work, we are interested in the motion of millimetre to sub-millimetre particles. The smaller the particle diameter the higher is the influence of the SRP (higher values of $\beta$ as can be inferred from \cref{eq:lightness_param}). As a consequence, we focus our attention on the anti-solar direction, where the L$_2$ point lies, and we neglect the point L$_1$, which moves too far away from the asteroid.

\section{Ejecta model}  \label{sec:ejecta_model}
\begin{figure}[htb!]
	\centering
	\includegraphics[width=2.6in]{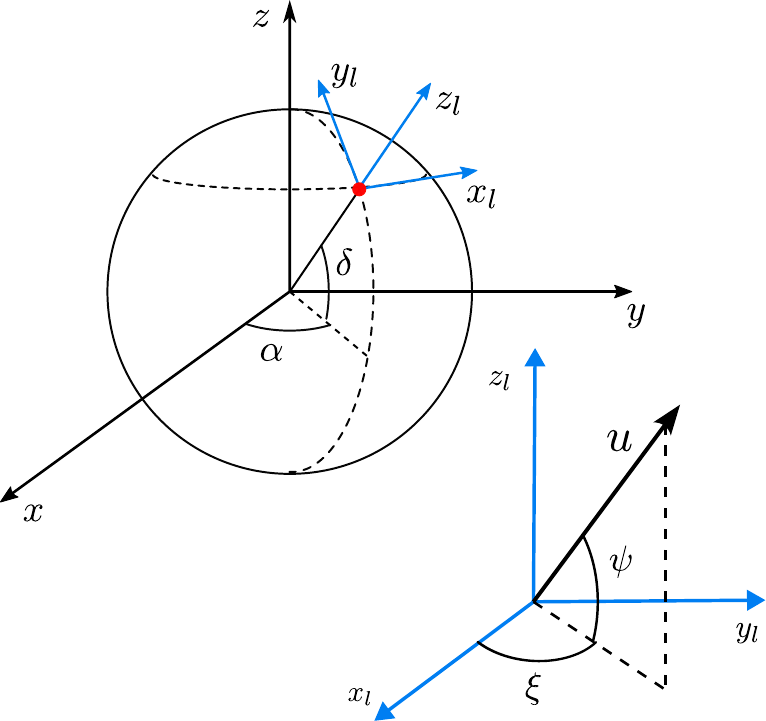}
	\caption{Synodic reference frame centred in the asteroid ($x,y,z$). The local horizontal frame ($x_l,y_l,z_l$) is tangent to the asteroid surface at the ejection location (red dot) identified by the right ascension, $\alpha$, and declination, $\delta$. The in-plane and out-of-plane ejection angles ($\xi$, $\psi$) identify the direction of the ejection speed, $u$, in the local horizontal frame.}
	\label{fig:ejecta_coords}
\end{figure}

This section describes the ejecta model used throughout this work. The ejecta model defines the characteristics of the ejected particles in terms of size, speed, and launch direction as they are generated by an impact on the asteroid surface. In this work, we consider hypervelocity impacts such as the one of the Hayabusa2 mission \citep{arakawa2020artificial,saiki2013small}, which can be generated using a small kinetic impactor carried by the spacecraft. The ejecta model we use is defined by a density function as follows \citep{sachse2015correlation}:

\begin{equation} \label{eq:ejecta_dist}
	\phi (s, u, \xi, \psi) = A s^{-1-\bar{\alpha}} u^{-1-\bar{\gamma}} f(\xi) g(\psi)
\end{equation}

where $s$ is the particle radius, $u$ the ejection velocity, $\xi$ the in-plane ejection angle, and $\psi$ the out-of-plane ejection angle (measured with respect to the local horizontal frame centred at the impact location) (\cref{fig:ejecta_coords}). 

The exponents $\bar{\alpha}$ and $\bar{\gamma}$ regulate the slope of the distribution, while $A$ is a scaling constant used for mass conservation. The model is a combination of uncorrelated distributions: the size, velocity and launch direction distributions are independent from each other \citep{sachse2015correlation,trisolini2021ejecta,trisolini2022scitech}. We assume the ejection of the particle can be uniform within a spherical sector so that the distribution functions for the launch directions can be expressed as:

\begin{equation} \label{eq:xi_dist}
	f(\xi) = \begin{cases}
		\frac{1}{\xi_{\rm max} - \xi_{\rm min}} \quad &\mathrm{if} \: \xi_{\rm min} \leq \xi \leq \xi_{\rm max} \\
		0 &\mathrm{elsewhere}
	\end{cases}
\end{equation}

\begin{equation} \label{eq:psi_dist}
	g(\psi) = \begin{cases}
		\frac{\cos{\psi}}{\sin{\psi_{\rm max}} - \sin{\psi_{\rm min}}} \quad &\mathrm{if} \: \psi_{\rm min} \leq \psi \leq \psi_{\rm max} \\
		0 &\mathrm{elsewhere}
	\end{cases}
\end{equation}

If we consider an impact perpendicular to the asteroid's surface, we can expect an axisymmetric ejecta cloud so that $\xi$ ranges from 0 to 360 degrees. Experiments show that the out-of-plane component of the ejection angle, $\psi$, usually ranges between 25 to 65 degrees \citep{richardson2007ballistics}. In this work, this range is adopted. The complete definition of the distribution function (\cref{eq:ejecta_dist}) requires the specification of additional parameters such as the minimum and maximum particle size. For this work, the minimum particle diameter is 10 $\mu$m, while the maximum is 10 cm. The selection of the particle size range is based on the work in \cite{yu2017ejecta,yu2018ejecta}, where the impact on the Didymos binary system is studied. Additionally, the minimum and maximum ejection speed must be specified. To do so, we can rely on experimental correlation for impacts and cratering events \citep{holsapple2012momentum,holsapple2007crater,housen2011ejecta}. The ejection velocity can be expressed as a function of the impactor and target properties as follows:

\begin{equation}  \label{eq:ejecta_speed}
	\frac{u}{U} = C_1 \left[ \frac{x}{a} \; \left( \frac{\rho}{\delta} \right)^{\nu} \right]^{-1/\mu}
\end{equation}

where $x$ is the radial distance from the centre of the crater, $U$ is the impactor speed, $a$ is the radius of the impactor, $\rho$ the target (asteroid) density, $\delta$ the impactor density, and $C_1$, $\mu$, and $\nu$ are constants that depend on the target material. \cref{tab:materials} shows the coefficients for the materials used in this study. The considered materials are \emph{Sand}, \emph{Weakly Cemented Basalt} (WCB), and \emph{Sand and Fly Ashes} (SFA), which are typical choices for modelling asteroid soils \citep{yu2018ejecta,richardson2007ballistics}. The first material is representative of very loose soil with zero equivalent strength, the second of low-strength material, and the third is representative of weakly cohesive soils, similar to regolith \citep{holsapple2012momentum}.

\begin{table}[htb!]
	\centering
	\caption{\label{tab:materials} Material properties.}
	\begin{tabular}{lccc}
		\hline
		& Sand & WCB & SFA \\
		\hline
		\hline
		$\mu$ 		& 0.41 	& 0.46 	& 0.4 \\
		$C_1$ 		& 0.55 	& 0.18 	& 0.55 \\
		$k$			& 0.3 	& 0.3 	& 0.3 \\
		$n_1$		& 1.2 	& 1.2 	& 1.2 \\
		$n_2$		& 1.3 	& 1 	& 1 \\
		$Y$ (MPa) 	& 0 	& 0.45 	& 4x10$^{-3}$ \\
		$A$ (\cref{eq:ejecta_dist}) & 2 & 2.7 & 2.4 \\
		\hline
	\end{tabular}
\end{table}

By substituting the minimum value, $x_{\rm min} = n_1 a$ (with $n_1$ = 1.2 \citep{housen2011ejecta}), we get the maximum ejection speed, while substituting the maximum value, $x_{\rm max} = n_2 \cdot R$, we get the minimum ejection speed. The crater radius, $R$, can be computed as follows \citep{housen2011ejecta}:

\begin{equation}  \label{eq:crater}
	\begin{cases}
		R \left( \frac{\rho}{m} \right)^{1/3} = H_2 \left( \frac{\rho}{\delta} \right)^{\frac{1-3\nu}{3}} \left[ \frac{Y}{\rho U^2} \right]^{-\mu/2} \quad\quad &\mathrm{strength}\\
		R \left( \frac{\rho}{m} \right)^{1/3} = H_2 \left( \frac{\rho}{\delta} \right)^{\frac{2+\mu-6\nu}{3(2+\mu)}} \left[ \frac{g a}{U^2} \right]^{-\mu/(2+\mu)} & \mathrm{gravity}
	\end{cases}
\end{equation}

where the first expression is for an impact in the strength-dominated regime, while the second for a gravity-dominated regime. The parameter $Y$ identifies the target strength, $g$ the gravitational acceleration at the target surface, and $m$ the impactor mass. With respect to the materials used in this study, for Sand material, the impact is gravity dominated, while for the other two types of materials, i.e., WCB and SFA, the crater formation is strength dominated \cref{eq:crater}.
From a comparison between the ejecta distribution of \cref{eq:ejecta_dist} and the expression derived by \cite{housen2011ejecta}, we can infer the expression for $\bar{\gamma} = 3 \mu$. Therefore, $\bar{\gamma}$ depends only on the target material. The coefficient $A$ can be obtained solving for the mass conservation as follows:

\begin{equation}
	\int_{\rm min}^{\rm max} \phi (s, u, \xi, \psi) ds du d\xi d\psi = M
\end{equation}

where \emph{min} and \emph{max} identify the minimum and maximum of the considered parameters, i.e., the particle size and velocity, and the in-plane and out-of-plane components of the ejection angles. The mass of the crater, $M$, can be computed as follows \citep{housen2011ejecta}:

\begin{equation}
	M = k \rho \left[ x^3 - (n_1 a)^3 \right]
\end{equation}

where $k$ is again a constant depending on the target material, which can be derived from experimental campaigns \cref{tab:materials}.

\begin{figure}[htb!]
	\centering
	\includegraphics[width=2.6in]{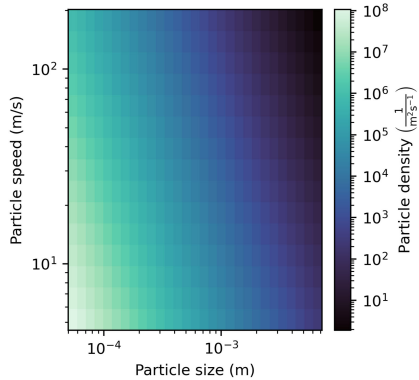}
	\caption{Example of ejecta model distribution obtained with \cref{eq:ejecta_dist}. Impact with a speed $U=2$ \si{\kilo\meter\per\second}, impactor diameter of 15 \si{\centi\meter}, and impactor mass of 2 \si{\kilo\gram} onto a sand-like target with equivalent strength $Y=0$ \si{\mega\pascal}.}
	\label{fig:ejecta_dist}
\end{figure}

An example of a 2D ejecta distribution in particle size and velocity is given in \cref{fig:ejecta_dist}. The main features of the ejecta model can be observed: the particle density increases rapidly (it is a log-log plot) with decreasing ejection speed and particle size so that after an impact event the majority of the particles will have small size and velocity.

\section{Collection strategies definition and target evaluation}  \label{sec:collection_strategies}

This section is dedicated to the description of the two possible collection strategies we have identified and of the procedure used to evaluate target asteroids \citep{trisolini2021ejecta}. The objective of the target evaluation procedure is to have a preliminary understanding of the possibility to collect particles in orbit, as a function of the macroscopic characteristics of an asteroid. We have identified as relevant characteristics the asteroid size, density, material, and equivalent soil strength. These parameters will be the subjects of a parametric analysis in \cref{sec:results}, whose aim is to understand if an in-orbit particle collection mission can benefit from specific asteroid characteristics.
In this work, the target selection evaluation is connected to the potential availability of particle for collection. Specifically, we are interested in particle size in the millimetre and sub-millimetre ranges. This choice is directly connected to the ejecta formation mechanisms, which generates higher number of particles at smaller sizes (\cref{fig:ejecta_dist}). Therefore, targeting smaller particles increases the chance for in-orbit collection to be successful.
As the particles of interest are small in size, their dynamics is strongly influenced by the solar radiation pressure. Smaller particles in most cases will be "swept" by the SRP towards the anti-solar direction. Larger particles can also escape the system, re-impact with the asteroid after one or more close passages, and, in some cases remain trapped in quasi-stable orbits for a long time \citep{pinto2020,scheeres2002fate}. Following these considerations, two collection strategies have been identified, which are described in \cref{subsec:antisolar} and \cref{subsec:orbiting}. The first strategy focuses on the anti-solar region after the asteroid as in this area it will be possible to collect some of the particles that will be swept by the SRP acceleration. The second strategy, instead, focuses on the region in the neighbourhood of the asteroid to understand the possibility to collect the particles that either re-impact or keep orbiting the asteroid for longer time. In \cref{subsubsec:antisolar_target} and \cref{subsubsec:orbiting_target} we will describe the target evaluation procedure for both these strategies to understand if different asteroid characteristics can favour a specific strategy.

The collection strategies and the target selection evaluation described in this section both focus on the dynamical behaviour of the particles as they move around the asteroid. No operational constraints have been directly analysed in this work, concerning in-orbit particle collection missions. For example, for the collection strategy in the anti-solar direction (\cref{subsec:antisolar}) may be more difficult to manoeuvre the spacecraft and eclipses will need to be considered. Nonetheless, at this stage, we are assuming that the proposed collection strategies can be carried out, so that we can analyse their potential. The decision to neglect operational constraints stems from the preliminary nature of the presented work. Once identified the families of asteroids with the highest potential for collection, more refined in-orbit collection analyses can be carried out, which will also include operational constraints. 

\subsection{Collection strategy in the anti-solar direction}  \label{subsec:antisolar}
\begin{figure}[htb!]
	\centering
	\includegraphics[width=2.75in]{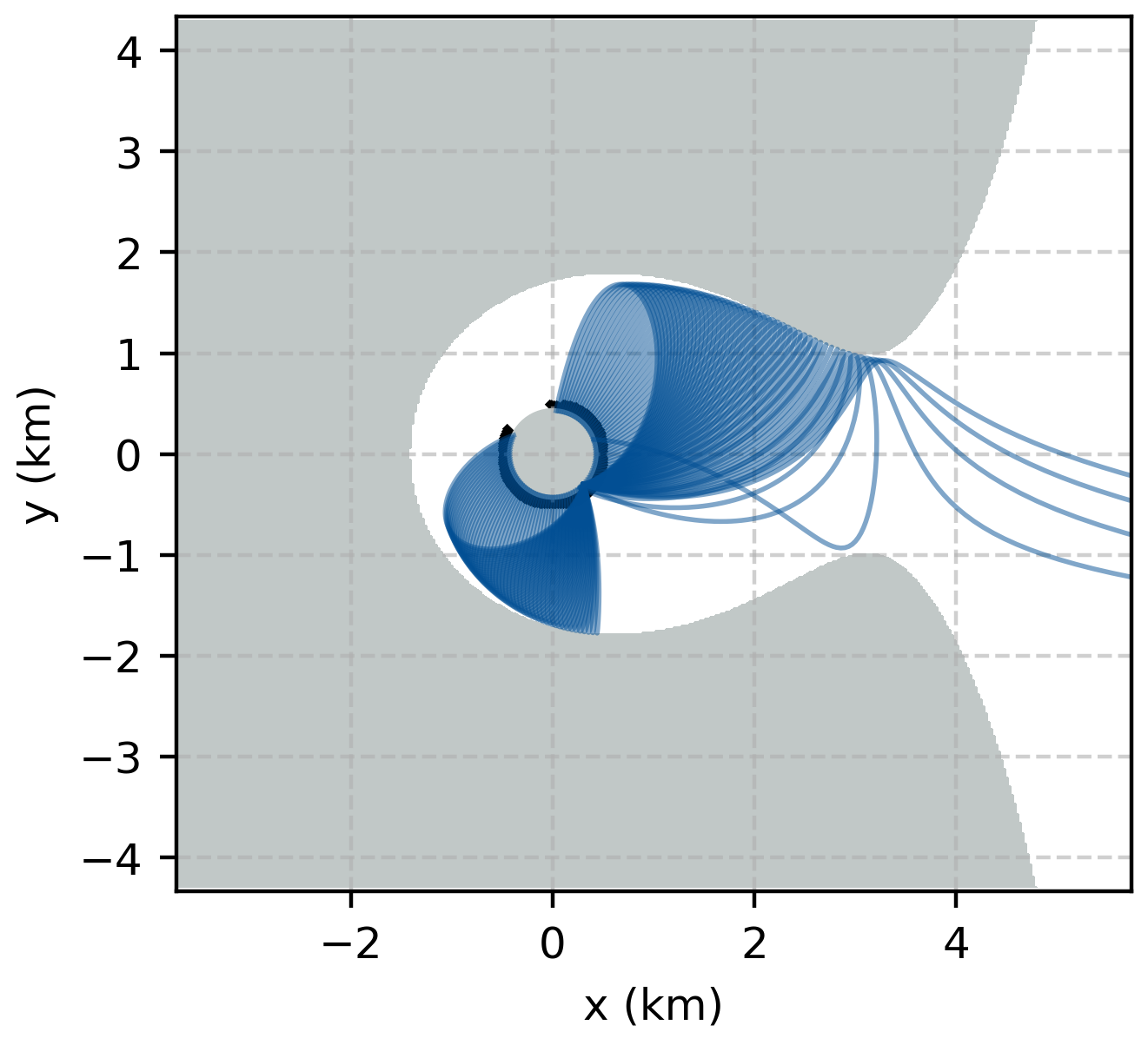}
	\caption{Zero-Velocity Curve, forbidden region (grey area), and example trajectories for a 1 mm particle having an energy level $C \leq C_2$.}
	\label{fig:zvc_trajectories}
\end{figure}

This section describes a collection strategy in which the spacecraft stations in the anti-solar region of the asteroids in order to collect the particles that, after the impact, will be carried in this direction by the effect of the SRP. For this strategy to be more effective, we can exploit the dynamical features of the CR3BP that is the presence of the L$_2$ Libration point along the anti-solar direction. In fact, we can compute the Jacobi constant (\cref{eq:jacobi}) in correspondence of the libration point L$_2$ setting the velocity to zero. We identify this Jacobi constant with $C_2$; a particle with an energy level $C \geq C_2$ will maintain a bounded motion around the asteroid because the zero-velocity curve is closed. On the other hand a lower level of energy may allow the particle to escape as a bottleneck opens up around L$_2$. It is possible to exploit this feature and target the collection of those particles that pass through the bottleneck at L$_2$. This strategy has thus the possible advantage to limit the collection operations to a more confined and predictable region. \cref{fig:zvc_trajectories} shows an example of zero-velocity curve together with a set of trajectories representative of an ejection condition with an energy level $C \leq C_2$, for a particle of 1 mm in size. Given the presence of the bottleneck at L$_2$, some of the particles can escape the system.

\begin{figure}[htb!]
	\centering
	\includegraphics[width=2.75in]{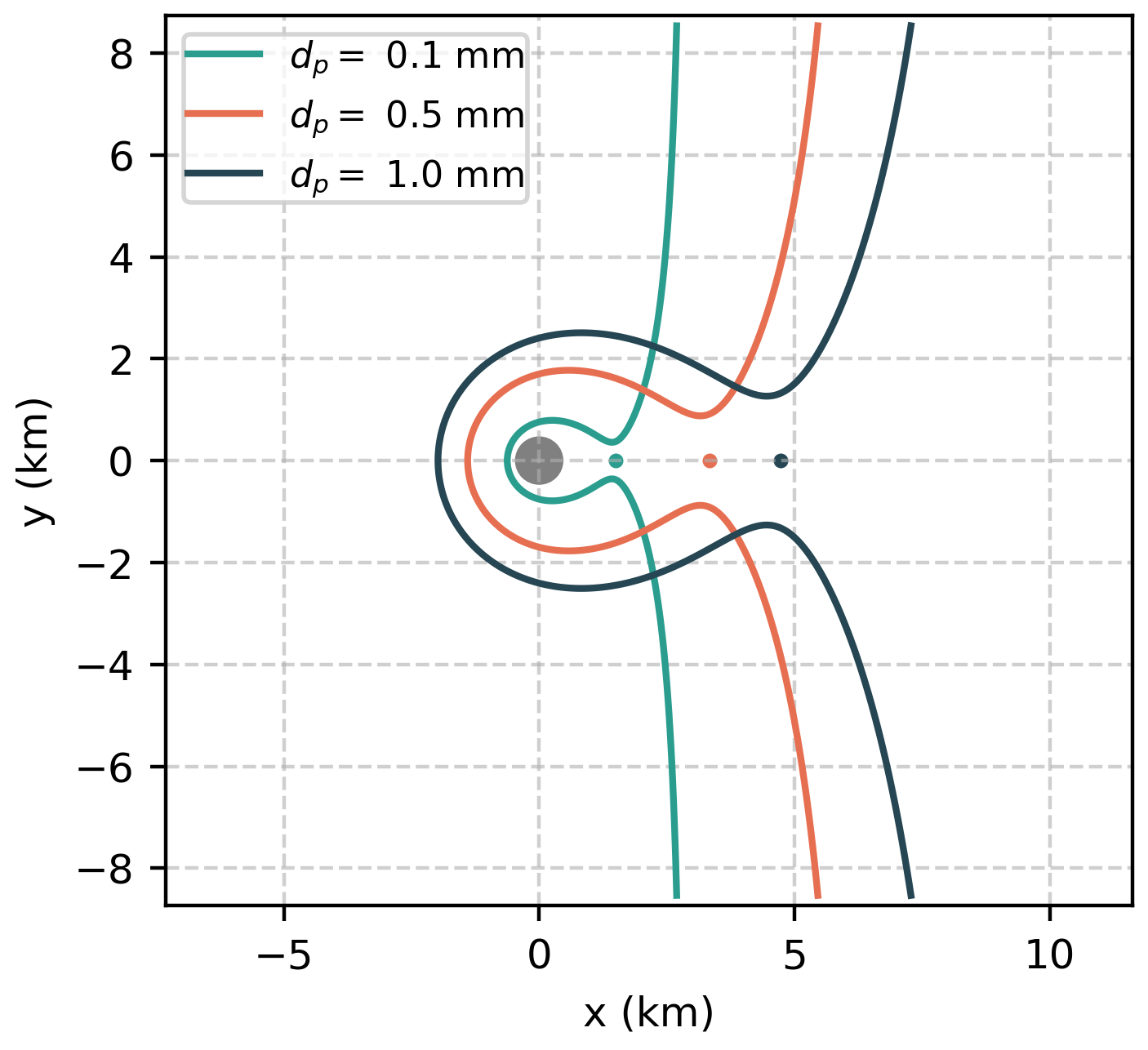}
	\caption{Zero-Velocity Curves and L$_2$ point location as a function of the particle diameter, with an energy level $C \leq C_2$, for an asteroid with 500 m radius.}
	\label{fig:zvc_dp}
\end{figure}

As mentioned in \cref{sec:dynamics}, the L$_2$ is an equilibrium point whose position depends on the magnitude of the SRP acceleration and, thus, from the particle diameter. Therefore, also the position and size of the bottleneck changes with the particle size. This behaviour is comparable to a mass spectrometer so that the collection of particles with different sizes may be performed controlling the distance from the asteroid along the x-axis of the synodic frame. \cref{fig:zvc_dp} shows three ZVC (the point represents the location of L$_2$) for three different particle diameters (0.1 mm, 0.5 mm, and 1 mm). We can observe that the position of L$_2$ and its associated gap moves closer to the asteroid as the particle size decreases. Given the link to the characteristics of the L$_2$ point, we refer to the strategy as the \emph{anti-solar} or L$_2$ collection strategy.

\subsubsection{Target evaluation}  \label{subsubsec:antisolar_target}
This section describes the target evaluation procedure for the anti-solar collection strategy. The objective of the target evaluation is to assess the potential of an asteroid to be suitable for the collection following the considered strategy, and to give the possibility to compare different target asteroids. We measure this potential using a Figure of Merit (FOM) as a proxy. In this work, the FOM is directly connected to an estimate of the number of particles that can be collected. After fixing the characteristics of the asteroid (size and density) and the material type and strength level, the procedure for the computation of the FOM is the following:

\begin{itemize}
	\item[1.] Select a \emph{test particle} of 1 mm in diameter. \\ \\
	As mentioned in \cref{sec:collection_strategies}, we are interested in millimetre and sub-millimetre particle sizes. Therefore, we use the 1 mm particle size as representative of the L$_2$ collection strategy. The concept of using a test particle as representative of the fate of an impact event has also been adopted in \cite{yu2017ejecta} for the Didymos binary system. 
	\item[2.] Compute the ejection velocity ($u_{\rm ej}$) such that a “small energy gap” opens up at L$_2$, as follows \citep{latino2019thesis,latino2019iac}: \\ \\
	$u_{\rm ej} = u_{C_2} + \epsilon \left(u_{esc} - u_{C_2} \right)$ \\ \\
	where $u_{\rm esc}$ is the escape velocity from the asteroid, $u_{C_2}$ is the escape velocity associated to an energy $C_2$, and $\epsilon$ is a correction coefficient that is used to slightly increase the energy of the system. Following \cite{latino2019thesis}, the $\epsilon$ coefficient is assumed equal to 0.025.
	\item[3.] Propagate a set of trajectories. \\ \\
	Assuming a 2D CR3BP, we perform an ejection simulation from a point on the asteroid surface every 10$^\circ$. For each ejection location, a set of trajectories is simulated for a range of ejection angles, with an interval of 5$^\circ$ between 25$^\circ$ and 65$^\circ$.
	\begin{figure}[htb!]
		\centering
		\includegraphics[width=2.6in]{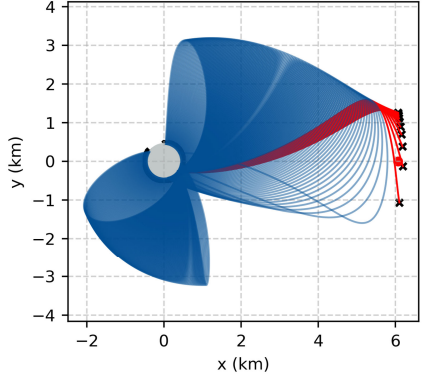}
		\caption{Examples of trajectories generated by a kinetic impactor. In red the trajectories passing via di L$_2$ gap.}
		\label{fig:traj_example}
	\end{figure}
	\item[4.] Estimate number of particles using ejecta distribution \\ \\
	To compute a Figure of Merit (FOM), we combine the simulations of Point 3 with the ejecta distribution of \cref{eq:ejecta_dist}. Since we assume that the ejected particles are uniformly ejected in the ejection direction, we can focus on the part of the distribution that is function of the particle size and speed, $\varphi(s, u)$. The ejecta distribution represents the particle density as a function of the initial ejection conditions. Therefore, by integrating it within a range of particle size and ejection speed, we can estimate the number of particles. To do so for the L$_2$ strategy, we take the test particle size, $s$. For each $k$-th bin in ejection location, we evaluate the ejection speed, $u_k$, as in Point 2. We estimate the number of particles associated to this combination of particle size and speed by integrating the ejecta distribution in a neighbourhood of $s$ and $u_k$:
	\begin{equation}  \label{eq:np_l2}
		n_{p,k} = \int_{s - \epsilon_s}^{s + \epsilon_s} \int_{u_k - \epsilon_u}^{u + \epsilon_u} \varphi(s, u) ds du
	\end{equation}
	where $\epsilon_s$ and $\epsilon_u$ define the neighbourhood of integration of the distribution. Since the conditions are specific, we select a very small neighbourhood that is $\epsilon_s = 1$ \si{\micro\meter} and $\epsilon_u = 0.1$ \si{\centi\meter\per\second}. For each ejection location, we have generated trajectories at different ejection angles and we need to take into account their contribution, because depending on the ejection angle, some particles will pass through the L$_2$ gap and some will not. Since the distribution is considered uniform in the ejection angle, for each ejection location the particle number $n_{p,k}$ is scaled with the fraction of trajectories passing through the L$_2$ gap (e.g., red lines in \cref{fig:traj_example}). The FOM is then the average over all the ejection locations as follows:
	\begin{equation}  \label{eq:fom_l2}
		\mathrm{FOM}_{L_2} = \log_{10} \left[ \frac{1}{M} \sum_{k=1}^{M} \frac{n_{L_2, k}}{N} \cdot n_{p, k} \right]
	\end{equation}
	where $M$ is the number of ejection locations, $N$ the total number of trajectories propagated from each ejection location and $n_{L_2,i}$ the number of trajectories passing through L$_2$ for each ejection location. Where we have introduced the logarithm for a better representation and comparison of the results.
\end{itemize}

Following the previously defined procedure it is possible to have an estimate of the average number of test particles available for collection in the anti-solar direction. As described in point 3 and 4, we perform an average over several ejection locations, instead of selecting the location with the highest number of potential particles. This choice derives from the fact that the selection of the impact location may depend on other constraints such as the mission operations or the texture of the soil. Therefore, an average value is a more robust metric.

\subsubsection{Preliminary operational considerations}
\label{subsubsec:operations_l2}
Despite the paper is focused on the assessment of possible targets via a simplified analysis, it is useful to provide a preliminary picture of the possible operations for the collection along the anti-solar direction. This type of collection, relies on the effect of SRP and on its "sweeping" effect on the small ejected particles. As such, the particles must be collected as they move away from the asteroid, and before they completely leave the system. Therefore, the operational time for the collection strategy will range from few hours to a about a day, depending on the asteroid size and its distance from the Sun. Depending on the objective of the mission (i.e., if specifics particle sizes are targeted) and the fuel limitations, the spacecraft can hover in a fixed location, "waiting" for the incoming particles or move along a trajectory trying to maximise the flux of collected particles.

\subsection{Collection strategy in the neighbourhood of the asteroid}  \label{subsec:orbiting}
\begin{figure}[htb!]
	\centering
	\begin{subfigure}[t]{0.32\textwidth}
		\centering
		\includegraphics[width=\textwidth]{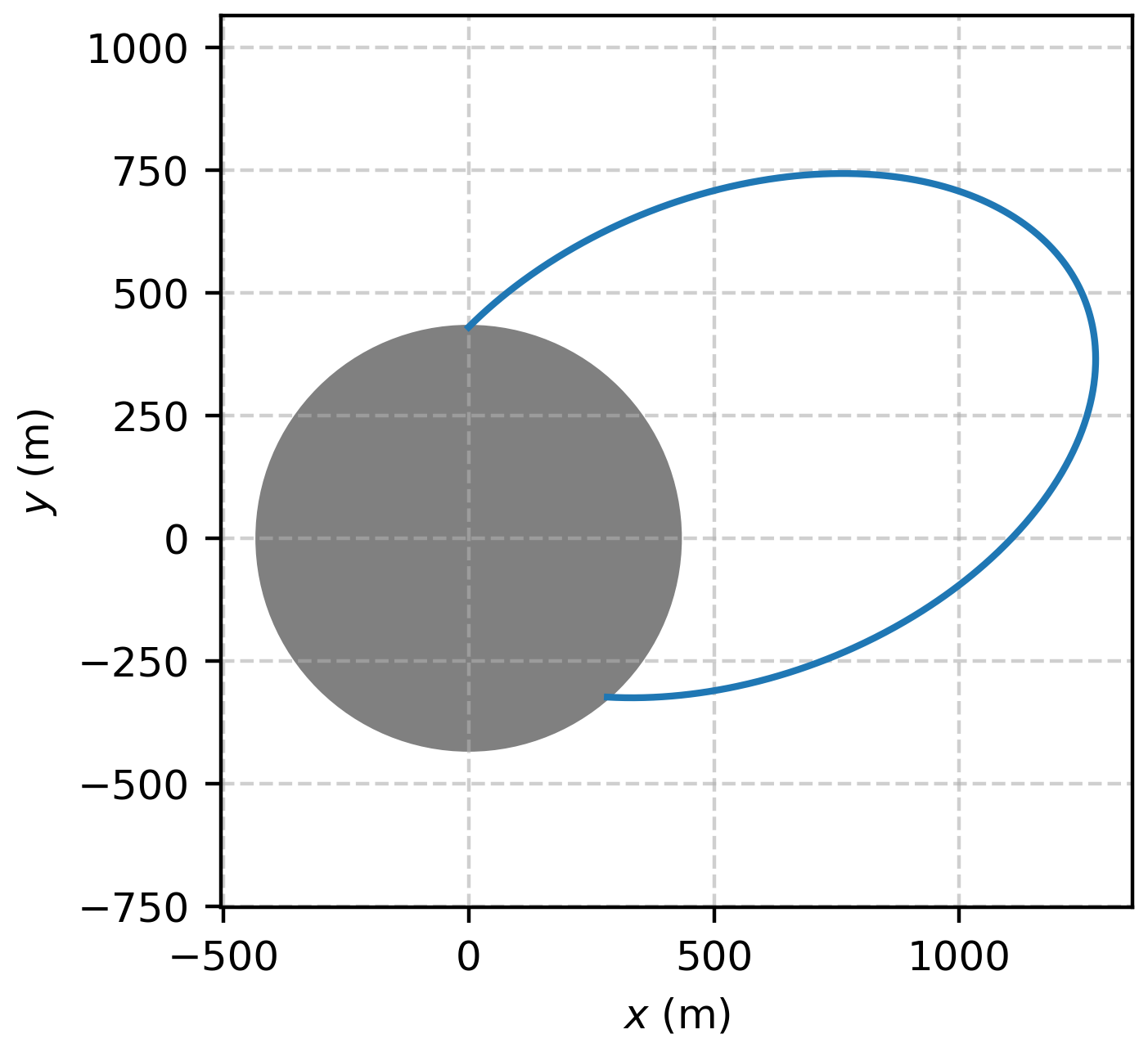}
		\caption{}
		\label{fig:orbit_parabolic}
	\end{subfigure}
	\hfill
	\begin{subfigure}[t]{0.32\textwidth}
		\centering
		\includegraphics[width=\textwidth]{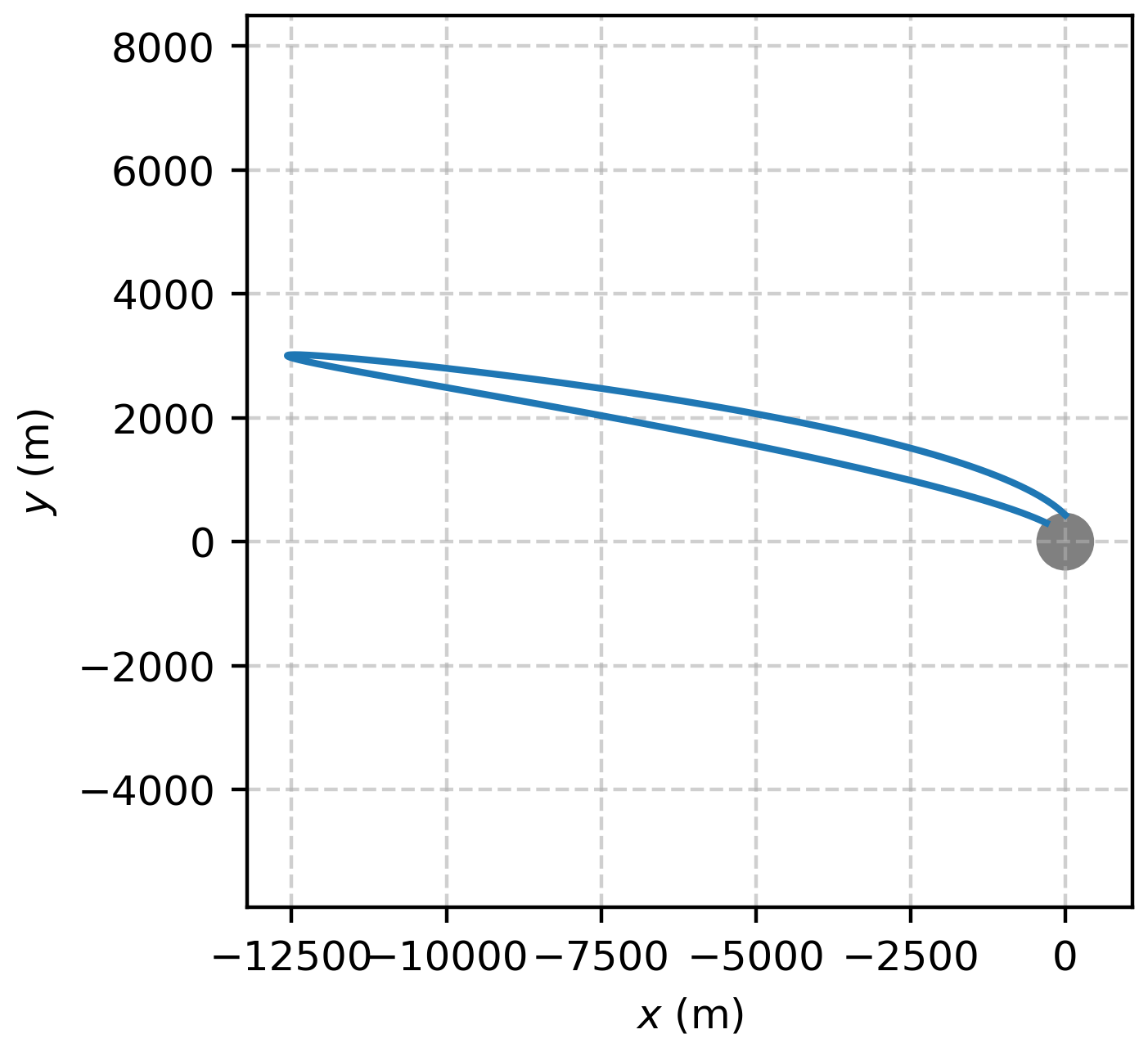}
		\caption{}
		\label{fig:orbit_impact_srp}
	\end{subfigure}
	\hfill
	\begin{subfigure}[t]{0.32\textwidth}
		\centering
		\includegraphics[width=\textwidth]{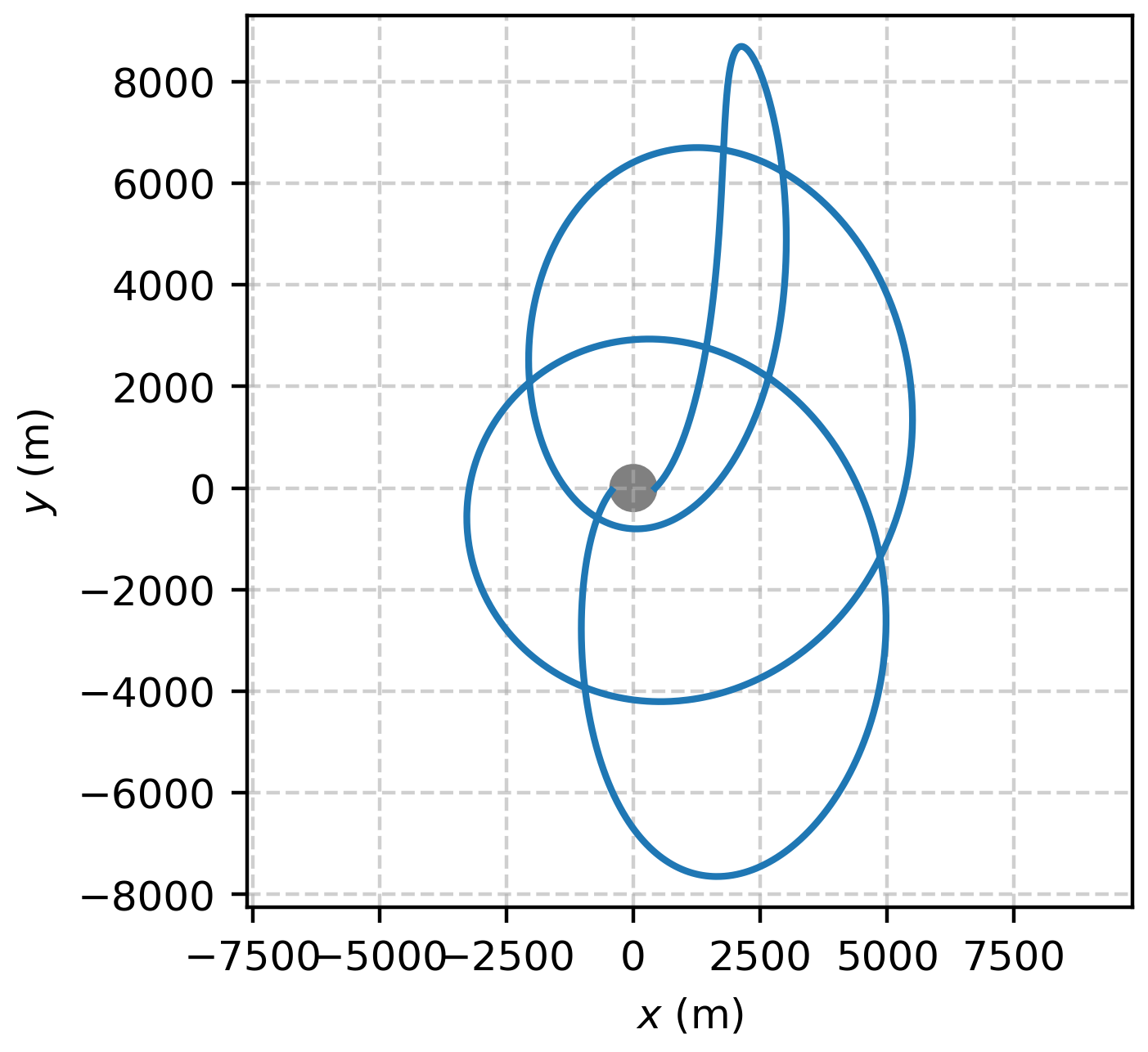}
		\caption{}
		\label{fig:orbit_quasistable}
	\end{subfigure}
	\caption{\label{fig:orbit_examples} Examples of particles' trajectories that can be exploited in the orbiting collection strategy. a) quasi-parabolic orbit b) SRP-driven re-impacting orbit c) orbit performing multiple close passages.}
\end{figure}

This second strategy is dedicated to the collection of those particles that, after an impact, will keep orbiting the asteroid for a sufficient amount of time. Instead, we neglect the particles quickly escaping the asteroid system. We will refer to this strategy as the \emph{orbiting} collection strategy. Different types of orbits can be of interest for the orbiting collection strategy: examples are the quasi-parabolic orbits (\cref{fig:orbit_parabolic}), the re-impacting orbits driven by the SRP (\cref{fig:orbit_impact_srp}), or orbits that perform multiple close approaches with the asteroid and will orbit for several days or more (\cref{fig:orbit_quasistable}). These and other types of orbital motions can all lead to particles that can be collected in a neighbourhood of an asteroid.

This strategy has the advantage of having less constraints on the initial conditions of the ejecta, with respect to the L$_2$ strategy. However, the particles can occupy a vast space around the asteroid, thus being less localised. Therefore, a more refined “collection trajectory” would be required for the spacecraft for an effective in-orbit collection. Depending on what type of particles' trajectories we target, different types of collection trajectories may be selected. For example, a hovering strategy at a specific distance from the asteroid may be beneficial to collect those particles following almost parabolic trajectories.

\subsubsection{Target evaluation}  \label{subsubsec:orbiting_target}
For the orbiting collection strategy, we consider collectable all those particles that orbit the asteroid for a time greater than a minimum required time, before re-impacting the asteroid surface. Therefore, we are interested in particles whose trajectory performs multiple close passages with the asteroids and orbits for several days and weeks, but also in those particles that have quasi-parabolic trajectories. However, we do not consider particles that escape the system. Similar to \cref{subsubsec:antisolar_target}, we introduce a Figure of Merit, FOM$_{\rm orb}$, that measures the particles available for collection. As in \cref{subsubsec:antisolar_target}, the target evaluation depends on the asteroid size, density, material type and strength. The procedure for the target evaluation is the following:

\begin{itemize}
	\item[1.] Specify the size range for the collectable particles: 0.1 – 2 mm in this study. \\ \\
	Also in this case, the size range is selected to focus on particles of millimetre and sub-millimetre scales.
	\item[2.] Select a range of ejection velocities. A simplified Keplerian motion is used to compute the minimum velocity, $u_{\rm min}$, by fixing the minimum time before re-impact ($T_{\rm min}$). The minimum speed is then computed as follows:
	
	\begin{align*}
		a_{T_{\rm min}} &= \left[\mu_A \left( \frac{T_{\rm min}}{2 \pi}\right)^2\right]^{1/3} \\
		v_{\rm min} &= \sqrt{ \frac{2 \mu_A}{R_A} - \frac{\mu_A}{a_{T_{\rm min}}}}
	\end{align*}
	
	where $a_{T_{\rm min}}$ is the semi-major axis of the particle orbit given the minimum time before re-impact, and $\mu_A$ and $R_A$ the gravitational parameter and radius of the asteroid, respectively. For the case in exam, the minimum time is fixed to 3 hours. This time limit has been selected to allow the spacecraft sufficient time to manoeuvre and get positioned in the spot dedicated for collection. The maximum velocity ($u_{\rm max}$) is set equal to the escape velocity of the asteroid. In this way, we focus on the particles that tend to remain in orbit around the asteroid.
	\item[3.] Propagate the trajectories. In this case, we use a grid in ejection location every 45$^\circ$, and a grid in ejection angle, $\psi$, every 5$^\circ$, between 25$^\circ$ and 65$^\circ$. We also use a grid in the particle size and speed, of 10 and 8 bins, respectively.
	\item[4.] Identify the percentage of trajectories still orbiting for a time greater than the minimum time. \\
	As the diameter influences the dynamical behaviour, different fractions of particles will be orbiting depending on their dimensions. Therefore, for each bin in particle size, we compute the fraction of particles ($w_k$) still orbiting after a time greater than $T_{\rm min}$:
	\begin{equation*}
		w_k = \frac{N(t_f > T_{\rm min})}{N},
	\end{equation*}
	where $N$ is the total number of trajectories propagated for each bin in size (this is the product of the number of bins in particle speed, ejection location, and ejection direction), and $N(t_f > T_{\rm min})$ is the number of these trajectories surviving for more than $T_{\rm min}$.
	\item[5.] Estimate number of particles and compute the Figure of Merit ($\mathrm{FOM}_{orb}$). \\
	The particle estimate is obtained using the ejecta distribution of \cref{eq:ejecta_dist}. In particular, we use a subset of the distribution in particle size and speed. We assume instead that the particles are uniformly ejected so that each trajectory in ejection direction can be weighted equally. As for Point 4, to estimate the number of particles we take into account the dependency in the particle size. For each $k$-bin in particle size, we first compute the minimum and maximum ejection speed ($u_{\rm min}^k$, $u_{\rm max}^k$) of the orbits satisfying the condition $t_f > T_{\rm min}$. Then, we integrate the distribution of \cref{eq:ejecta_dist}, $\varphi(s, u)$, inside the range defined by the bin in particle size and the minimum and maximum ejection speeds. The obtained number of particles is then scaled with the fraction computed in Point 4, to take into account that not all the trajectories satisfy the condition $t_f > T_{\rm min}$. The expression for the FOM$_{\rm orb}$ is as follows:
	
	\begin{equation}  \label{eq:fom_orb}
		\mathrm{FOM}_{orb} = \log_{10} \left[ \sum_{k=0}^{n_d} \left( \int_{s_k}^{s_{k+1}} \int_{u_{min}^k}^{u_{max}^k} w_k \cdot \varphi(s, u) ds du \right) \right]
	\end{equation}
	
	where $n_d$ is the number of bins in which the particle size range is subdivided.  
\end{itemize}

It is important to note that the selected Figure of Merit gives an optimistic estimate of the number of collectable particles; indeed, it can be interpreted as the maximum number of particles available for collections. The actual number of particles collected will then depend on the specific mission architecture and collection device. Nonetheless, the superiority of a target asteroid with respect to another can, in a preliminary analysis such as this one, be represented by the proposed FOM of \cref{eq:fom_orb}, as it represents the highest potential for collection for a given mission.

\subsubsection{Preliminary operational considerations}
\label{subsubsec:operations_orbiting}
As in \cref{subsubsec:operations_l2}, we provide here a brief outline of the preliminary operational considerations for the orbiting collection strategy. For this strategy, we can subdivide the possible operations in two phases. A first phase that can focus on particles re-impacting the asteroid before their first pericentre passage (e.g., \cref{fig:orbit_parabolic}). These particles follow quite predictable trajectories and their impact time can be estimated. Therefore, the spacecraft can be placed in a hovering position above the asteroid's surface in a specified location, waiting for the particles. This phase is quite rapid and can last between few hours and few tens of hours. In a second phase, the spacecraft can focus on particles staying for a longer time around the asteroid. These are the particles that perform multiple revolutions around the asteroids; therefore, this phase can last between days and weeks. In general, these particles will be larger in size and will be less abundant; hence, their collection will be more challenging. The chance of collection can be increased by bounding the motion of the spacecraft to specific regions where we expect a higher particle concentration, which depend on the impact location.

\section{Results}  \label{sec:results}


In this section, the procedures described in \cref{subsubsec:antisolar_target,subsubsec:orbiting_target} are utilised to assess the potential for in-orbit collection of target asteroids. As we want to identify the most suitable families of targets for such a collection scenario, it is necessary to evaluate a large combination of target characteristics, based on their density, dimensions, material type, and material strength. In fact, these parameters are the most influential in determining the properties of the ejecta plume and, consequently, the particle availability around the asteroids. It is important at this point to recall that the composition of most asteroids is still unknown, leading to a considerable uncertainty when it comes to predicting their density and soil strength. While the density can be connected to the spectral type of the asteroid \citep{carry2012density} to reduce such uncertainties, the soil type and strength is more difficult to predict. As these parameters have a strong influence on the outcome of the ejecta model, it is critical to perform a parametric analysis considering their effects.

To do so, the presented analysis, considers a combination of possible asteroids sizes and densities. Scanning through these combinations we can try to identify the more promising targets as a function of observable characteristics, by ranking them using the evaluation procedures of \cref{subsubsec:antisolar_target,subsubsec:orbiting_target}. Specifically, we vary the radius of the asteroid from 100 \si{\meter} to 15 \si{\kilo\meter}, while the asteroid density from 1 \si{\gram\per\centi\meter\cubed} to 5.3 \si{\gram\per\centi\meter\cubed}. The radius range derives from data in the NASA asteroid small body database, excluding small objects. The density ranges are derived from average densities of common asteroid spectral classes \citep{carry2012density}. The previously defined radius and density ranges are subdivided into a 25$\times$25 grid. Therefore the target evaluation procedure is repeated for 125 combinations, to obtain a map that shows the Figure of Merit as a function of the asteroid radius and density (e.g., \cref{fig:fom_plot_l2}). We chose to perform the target evaluation on a grid of radius and density instead of evaluating each asteroid in the NEA database because in this way we have a more general understanding of what characteristics we should look for to have a more favourable in-orbit collection. In addition, there is a total of more than 25 thousand NEAs catalogued so far, making it impractical to perform the simulations of \cref{subsubsec:antisolar_target,subsubsec:orbiting_target} for each one of them. Also, having a more generic map, allows us to evaluate newly discovered asteroids by placing them inside the map, given their macroscopic properties.

Because the target evaluation depends upon the material type and strength, a different map can also be obtained for different combination of materials and strengths. Specifically, in this work, we model the sand-like material with null strength (0 \si{\kilo\pascal}), the SFA material with two strength levels (1 and 4 \si{\kilo\pascal}), and the WCB material with four levels (1, 10, 50, and 500 \si{\kilo\pascal}). It is possible to observe that for the WCB and SFA materials we consider strength levels below the reference value reported in \cref{tab:materials}. This decision follows the consideration that asteroid soils are predicted to be weaker than the corresponding sample terrains used for ground experiments. \cite{richardson2007ballistics} found 10 \si{\kilo\pascal} to be a reasonable value for the soil strength of comet Temple1, the strength of lunar regolith is estimated between 1 and 3 kPa \citep{holsapple2007crater}, and the impact on asteroid Ryugu showed a very weak soil, with an impact almost fully gravity dominated \citep{arakawa2020artificial}. 

To fully define the ejecta model, the definition of the impactor characteristics is also required. In this work, we use a fixed impactor for all the simulations, whose properties are summarised in \cref{tab:impactor} and the direction of impact is assumed to be perpendicular to the asteroid's surface. The impactor has the same characteristics of Hayabusa2's Small Carry-On Impactor (SCI) \citep{tsuda2019hayabusa2,soldini2017assessing}. The variation of the impactor characteristics has only a limited influence on the minimum ejecta speed. Therefore, it has a limited influence on the fraction of particles available for collections, which is mainly driven by the low-speed portion of the ejecta distribution \cite{trisolini2021jaxaws}.

\begin{table}
	\centering
	\caption{Summary of the impactor properties \citep{tsuda2019hayabusa2,soldini2017assessing}. \label{tab:impactor}}
	\begin{tabular}{lcc}
		\hline
		Quantity & Symbol & Value \\
		\hline
		Speed (km/s) & $U$ & 2 \\
		Radius (m) & $a$ & 0.075 \\
		Density (g / cm$^3$) & $\delta$ & 8.9 \\
		Mass (kg) & $m$ & 2 \\
		\hline
	\end{tabular}
\end{table}

As described in \cref{sec:dynamics}, we model the target asteroids as spheres; therefore, the effect of irregular gravity field is neglected. This assumption follows from few considerations. The performed analysis is preliminary in nature and is dedicated to identify suitable characteristics for in-orbit sample collection that can allow us to select a potential target over another by comparing their FOM. This analysis is based on basic properties of the asteroids and the ones we consider most influential in the ejecta generation process. The effect of an irregular gravity field, statistically, has a smaller influence on the overall fate of the ejecta \citep{zanotti2019hypervelocity,matsumoto2011numerical,scheeres2000temporary}, particularly for smaller particles sizes such as the one we are interested. As the procedures of \cref{subsubsec:antisolar_target,subsubsec:orbiting_target} are interested in the overall fate of the ejecta, it was decided to neglect such a contribution. In addition, as we are not analysing each NEA individually, we cannot associate a specific irregular gravity field to the generic combinations of the radius-density grid we use. Also, information on the gravity field is not available for most of the asteroids in the NEAs database \citep{nasa_sbdb}.

A final assumption involved in the target selection analysis concerns the semi-major axis of the target asteroid orbit used for the computation of the map. The semi-major axis is assumed to be constant and equal to 1.755 AU for the different combinations of asteroid size, density, material, and strength. This value correspond to the average semi-major axis of the orbits of all NEAs in the NASA Small Bodies Database \citep{nasa_sbdb}. This choice results from the necessity to limit the number of simulations (and thus the computational time), while retaining the representativeness of the results. In fact, considering a variable semi-major axis for the asteroid would require an additional map in asteroid radius and density for each bin in semi-major axis. In addition, the majority of the NEAs (about 63\%) has a semi-major axis within one sigma (0.56 AU) of the mean value.

\subsection{$L_2$ strategy results}  \label{subsec:l2_results}
\begin{figure}[htb!]
	\centering
	\begin{subfigure}[t]{0.45\textwidth}
		\centering
		\includegraphics[width=\textwidth]{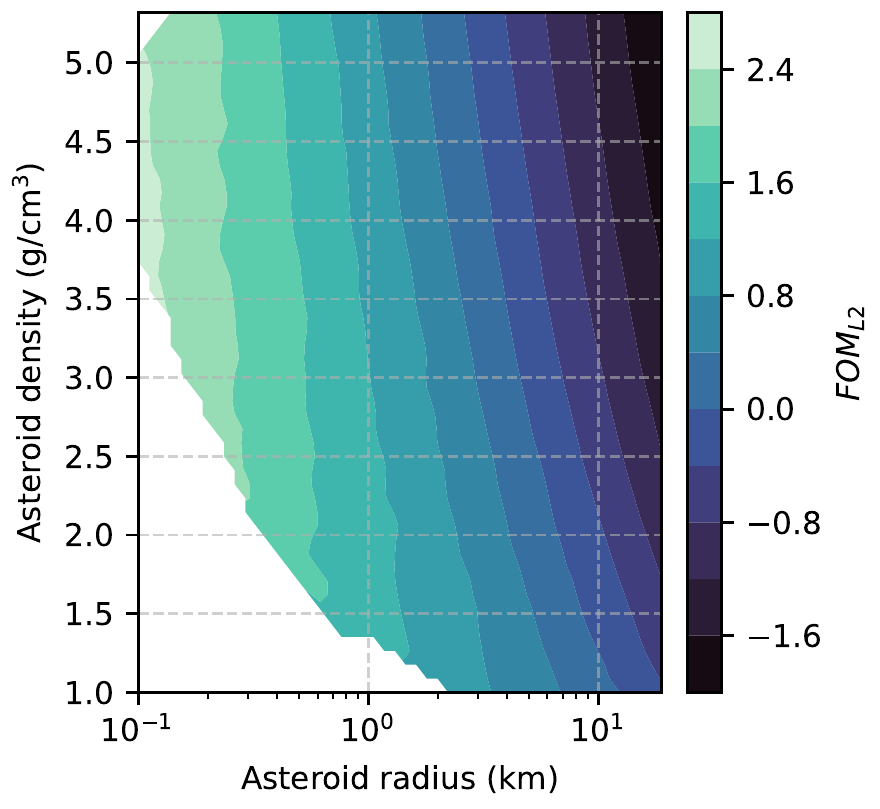}
		\caption{}
		\label{fig:fom_sand_l2}
	\end{subfigure}
	\hfill
	\begin{subfigure}[t]{0.45\textwidth}
		\centering
		\includegraphics[width=\textwidth]{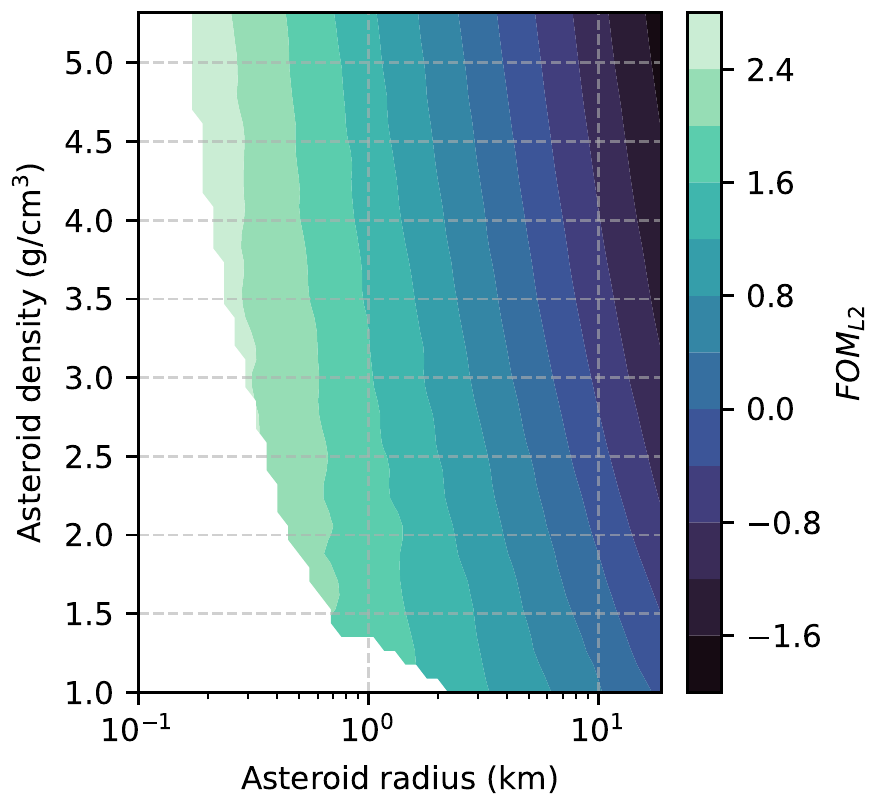}
		\caption{}
		\label{fig:fom_wcb_l2}
	\end{subfigure}
	\hfill
	\begin{subfigure}[t]{0.45\textwidth}
		\centering
		\includegraphics[width=\textwidth]{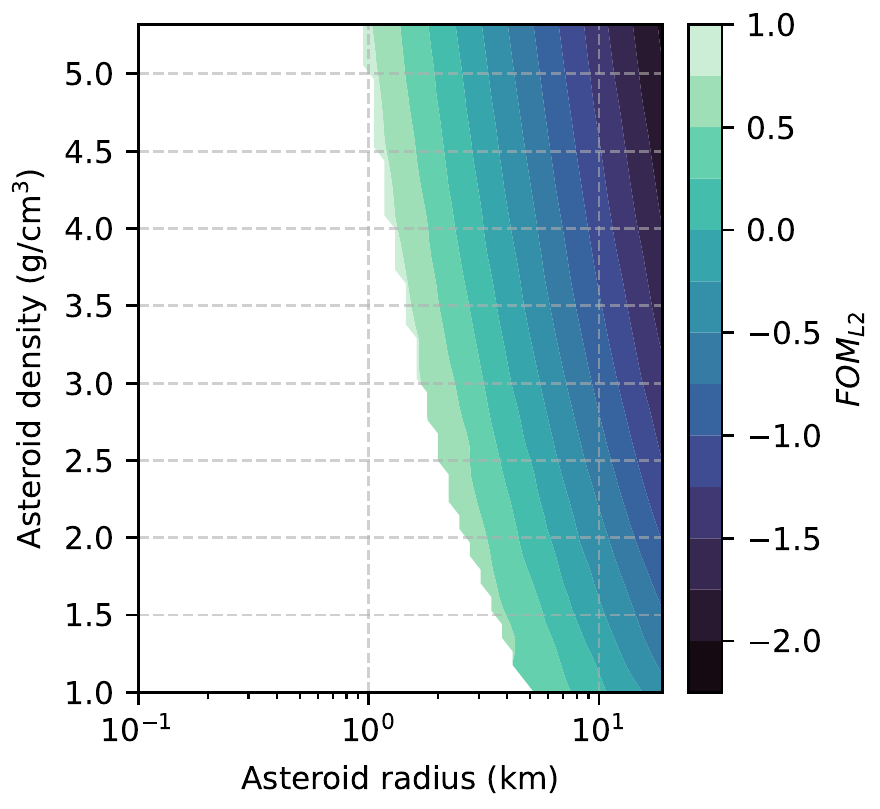}
		\caption{}
		\label{fig:fom_sfa_l2}
	\end{subfigure}
	\caption{\label{fig:fom_plot_l2} FOM as function of the target size and density for the $L_2$ collection strategy. (a) Sand-like material ($Y$ = 0 \si{\pascal}). (b) WCB ($Y$ = 1 \si{\kilo\pascal}). (c) SFA ($Y$ = 4 \si{\kilo\pascal})}
\end{figure}

This section presents the preliminary collectability analysis for the $L_2$ collection strategy (as described in \cref{subsec:antisolar}), comparing the results for three different materials. The first result of the study consists in the generation of maps of \emph{collectability scores}, representing the Figure of Merit ($\mathrm{FOM}_{L_2}$) as a function of the asteroid radius and density. \cref{fig:fom_plot_l2} shows examples of such maps for three different cases. A first case, \cref{fig:fom_sand_l2}, for which the FOM is evaluated for a sand-like material; a second case, \cref{fig:fom_wcb_l2}, for the WCB material assuming a strength of 1 \si{\kilo\pascal}; and a third case, \cref{fig:fom_sfa_l2}, for the SFA material with a strength of 4 \si{\kilo\pascal}. For a better representation we use a logarithmic scale for the asteroid radius; the FOM already represents the logarithm of the number of potentially collectable particles.

The first feature we can observe is the presence of infeasibility regions in all the maps, with a trend that shows increasing infeasibility regions as the strength of the material increases. This trend is commonly present for all the materials but the influence of the material strength can be more or less pronounced. The infeasibility regions describes those combinations of characteristics of a target asteroid for which none of the simulated trajectories passes through the $L_2$ gap, as defined in Point 2 of \cref{subsubsec:antisolar_target}. Another observable feature is the location of the best and worst collection cases. The best case il located at the upper left part of the feasible region, corresponding to small, high density target asteroids. Instead, the worst solution is situated at the upper right corner, corresponding to large, high density asteroids. Because in general the density and material strength of an asteroid can be uncertain, it would be preferable, for a more robust target selection, to stay at enough distance from the infeasibility region. Therefore, larger and denser asteroids would provide a safer option for collection as they are further away from the infeasibility region; however, they also have a lower ranking in terms of the Figure of Merit.
A final observation on the $L_2$ collection strategy concerns the values of the $\mathrm{FOM}_{L_2}$ index. Comparing the magnitude of the index between the different materials, we observe comparable values between the sand-like and WCB material, while the SFA material shows lower values, corresponding to a one order of magnitude difference for the number of particles. The overall magnitude has low overall values of the FOM, corresponding to tens and hundreds of particles. These are small values; however, it is important to recall that the index is estimated considering only a specific value of the test particle diameter and ejection speed. Therefore, it is used as a relative measure of the availability of particles at the $L_2$ gap.

\begin{figure}[htb!]
	\centering
	\includegraphics[width=3.4in]{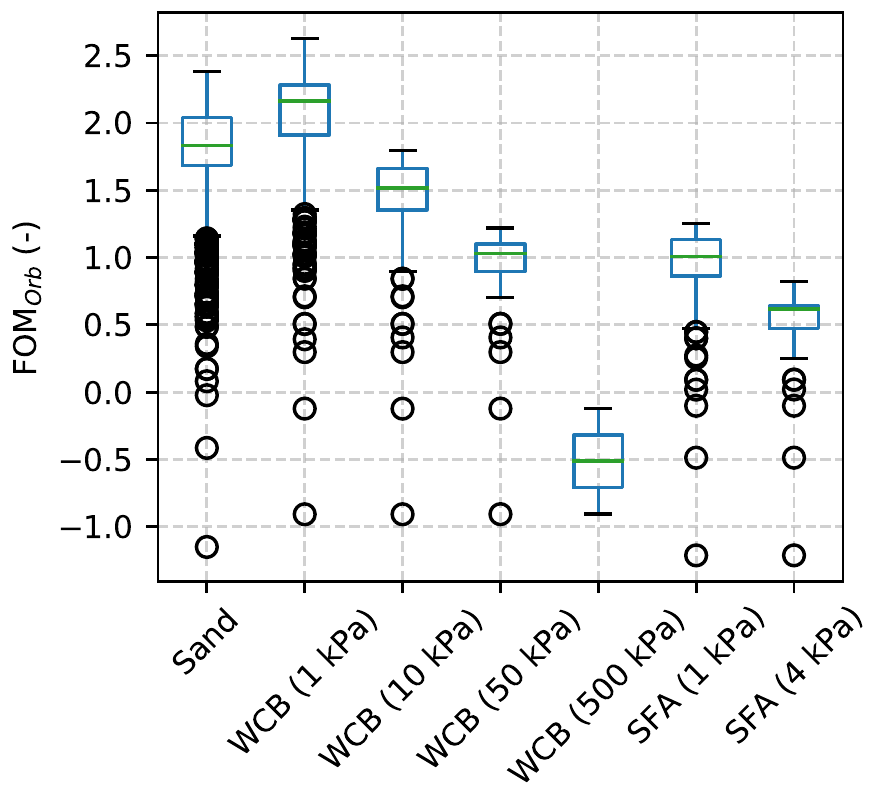}
	\caption{Boxplot of the $\mathrm{FOM}_{L_2}$ as function of the material type and strength.}
	\label{fig:boxplot_fom_l2}
\end{figure}

\bigbreak
An interesting application of the previously presented maps is the computation of the $\mathrm{FOM}_{L_2}$ index for the asteroids in the NASA Small Bodies Database \citep{nasa_sbdb} by just locating them on the maps as a function of their mean radius and density. Whenever the size was not available, it was estimated from the albedo and magnitude information\footnote{\url{https://cneos.jpl.nasa.gov/tools/ast_size_est.html}}. Similarly, the density has been estimated from the spectral class of the asteroid \cite{carry2012density} and, when not available, assuming an average density of 2.6 \si{\gram\per\centi\meter\cubed}.
If we apply this procedure, we can evaluate the Figure of Merit for the overall asteroid population and compare it for different material types and strengths. \cref{fig:boxplot_fom_l2} shows this comparison. In this plot, the green line represents the median value of the index, the box represents the first and third quartiles (it contains 50\% of the cases), while the upper and lower whiskers (black lines) represent the third and first quartiles, plus and minus 1.5 times the interquartile range (IRQ), respectively. The remaining dots can be considered outliers.

Interesting features can be observed from \cref{fig:boxplot_fom_l2}. First, the average Figure of Merit for a given material type decreases as the strength increases. In fact, we have a three orders of magnitude difference between the WCB case at 1 \si{\kilo\pascal} with respect to the one at 500 \si{\kilo\pascal}. This is mainly due to the high ejection speeds generated by high strength materials so that most of the ejecta rapidly escape the neighbourhood of the asteroid. Second, the index of the sand-like material is comparable to a low-strength WCB material, while SFA materials show low index values despite their low strength. These results show the influence of the material type on the possible outcome of an impact event. Therefore, to ensure the robustness of the target selection, it of paramount importance to perform analyses considering different combinations of material types and strengths.

\begin{figure}[htb!]
	\centering
	\includegraphics[width=3.4in]{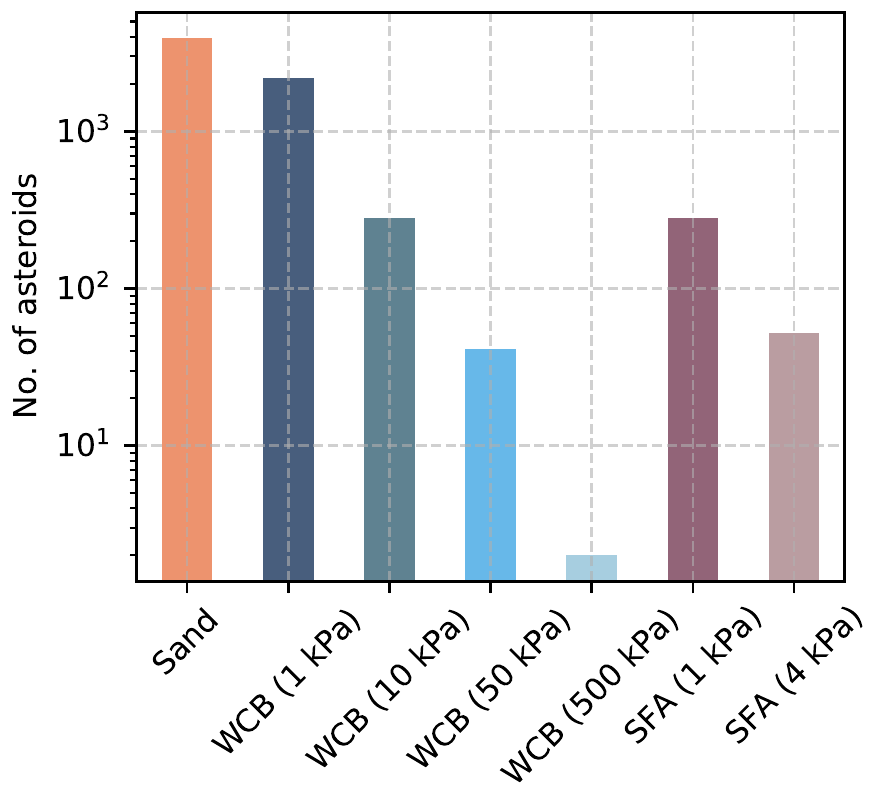}
	\caption{Number of identified targets for $L_2$ collection as function of target material. Test particle: 1 \si{\milli\meter} size.}
	\label{fig:barplot_count}
\end{figure}

Because the infeasibility region changes as a function of the material type and its strength, we have a different number of possible targets for the analysed cases. \cref{fig:barplot_count} shows the number of identified targets for the $L_2$ collection option for the different cases, assuming all the targets are modelled with one material-strength combination. It is interesting to observe that the number of identified targets follows the trend of the Figure of Merit, with some minor differences. The highest number of feasible targets (about 3900) is associated to the sand-like material, while the lowest number of targets (only 8) is associated to the high-strength WCB option.

\bigbreak
Because of the intrinsic uncertainty concerning the modelling and the characteristics of asteroid materials, this type of analysis allows the evaluation of the robustness of the $L_2$ particle collection scenario. In fact, the list of identified feasible targets can be selected from the more conservative higher-strength case, to create a shortlist of possible targets. Of course, the 500 \si{\kilo\pascal} WCB case would be extremely restrictive. However, recent missions such as Hayabusa2 and OSIRIS-Rex, and past missions such as Deep Impact have shown low strength properties of the explored asteroid's soil: asteroid Ryugu experienced a gravity-dominated crater formation when hit by Hayabusa2's small kinetic impactor, indicative of very low-strength soils; OSIRIS-Rex showed that asteroid Bennu has a terrain similar to Ryugu and the tag instrument penetrated the soil further than expected, indicating a soft terrain; \cite{richardson2007ballistics} estimated the strength of the soil of comet Temple1 to be around 10 \si{\kilo\pascal}. Given these considerations, it would be plausible to select an intermediate soil strength (between 10 and 50 \si{\kilo\pascal}) for a selection of possible targets. \cref{tab:rank_l2} shows a shortlist of the asteroids ranked from the lowest $\Delta v$ required to reach them. We compute the $\Delta v$ with a simplified two-impulse trajectory, following the procedure outlined in \cite{shoemaker1978earth}. The total $\Delta v$ is the sum of the contribution required to inject the spacecraft into a transfer trajectory from Low Earth Orbit and a second impulse needed to rendezvous with the asteroid. \cref{tab:rank_l2} collects some examples of possible targets, where we have restricted our research to all NEAs with an assigned name (adding unnamed asteroids would change part of the ranking) and we have kept only the asteroids having a valid index for at least four out of the seven different material-strength combinations. This last decision is excluding several target that only had collection possibilities for low-strength materials; however, we decided to present a more robust subset of possible targets. Among the table's entries, we highlighted in bold the three best asteroids in terms of Figure of Merit. Interestingly, we also observe two cases (Eros and Eric) with negative values of the index (i.e., less then one test particle passing through the $L_2$ gap), which are two very large asteroids (diameters greater than 10 \si{\kilo\meter}) and average densities (2.6 - 2.7 \si{\gram\per\centi\meter\cubed}) that are located in the right portion of the FOM maps (\cref{fig:fom_plot_l2}). We can also observe that all the possible targets require $\Delta v$ below 7 \si{\kilo\meter\per\second} and three of them below 6 \si{\kilo\meter\per\second}. Considering the $\Delta v$ estimates are conservative and do not take into account gravity assist manoeuvres, the identified targets are relatively affordable.

\begin{table}[htb!]
	\centering
	\caption{\label{tab:rank_l2} Target asteroid ranking for the $L_2$ collection strategy. Ranking from the lowest $\Delta v$.}
	\begin{tabular}{lccccccc}
		\hline
		Name &  Sand & WCB & WCB & WCB & SFA & SFA & $\Delta v$ \\
		 &   & 1 kPa & 10 kPa & 50 kPa & 1 kPa & 4 kPa & (km/s) \\
		\hline
		\textbf{Anteros} &   \textbf{1.22} &        \textbf{1.66} &         \textbf{1.64} &          - &          \textbf{1.11} &         - &       \textbf{5.11} \\
		Eros &  -0.41 &       -0.12 &        -0.12 &        -0.12 &        -0.49 &       -0.49 &       5.58 \\
		\textbf{Zephyr} &   \textbf{1.31} &        \textbf{1.76} &         \textbf{1.72} &          - &            \textbf{1.18} &         - &       \textbf{5.86} \\
		Geographos &   1.14 &        1.56 &         1.57 &          - &       1.05 &         - &       6.08 \\
		Ivar &   0.17 &        0.51 &         0.51 &         0.51 &         0.09 &        0.09 &       6.15 \\
		McAuliffe &   1.08 &        1.49 &         1.50 &          - &           0.99 &         - &       6.18 \\
		Seleucus &   0.97 &        1.38 &         1.38 &          - &         0.88 &         - &       6.20 \\
		Oze &   0.53 &        0.90 &         0.90 &         0.90 &         0.45 &        0.45 &       6.33 \\
		Toro &   0.96 &        1.35 &         1.35 &          - &          0.86 &         - &       6.43 \\
		Toutatis &   0.65 &        1.02 &         1.02 &         1.02 &        0.56 &        0.56 &       6.50 \\
		Melissabrucker &   0.56 &        0.92 &         0.92 &         0.92 &         0.47 &        0.47 &       6.60 \\
		Beltrovata &  1.22 &        1.66 &         1.64 &          - &       1.11 &         - &       6.66 \\
		\textbf{Cacus} &  \textbf{1.31} &        \textbf{1.76} &         \textbf{1.72} &          - &        \textbf{1.18} &         - &       \textbf{6.93} \\
		Alinda &  0.79 &        1.17 &         1.17 &          - &          0.70 &        0.69 &       6.96 \\
		\hline
	\end{tabular}
\end{table}

\subsection{Orbiting strategy results}  \label{subsec:orbit_results}
\begin{figure}[htb!]
	\centering
	\begin{subfigure}[t]{0.45\textwidth}
		\centering
		\includegraphics[width=\textwidth]{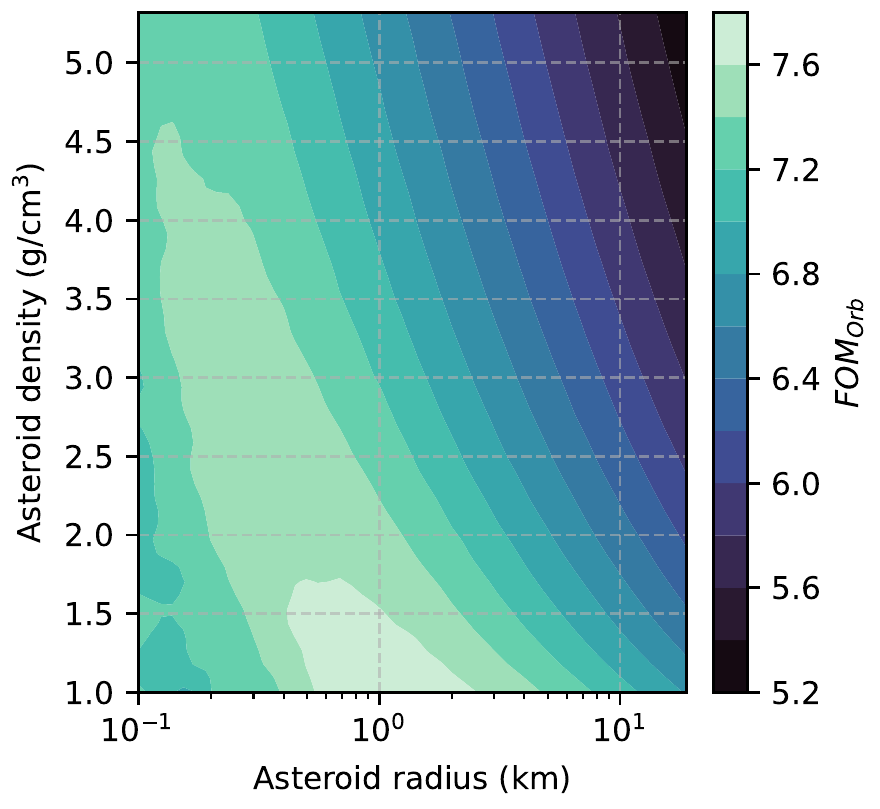}
		\caption{}
		\label{fig:fom_sand_orbit}
	\end{subfigure}
	\hfill
	\begin{subfigure}[t]{0.45\textwidth}
		\centering
		\includegraphics[width=\textwidth]{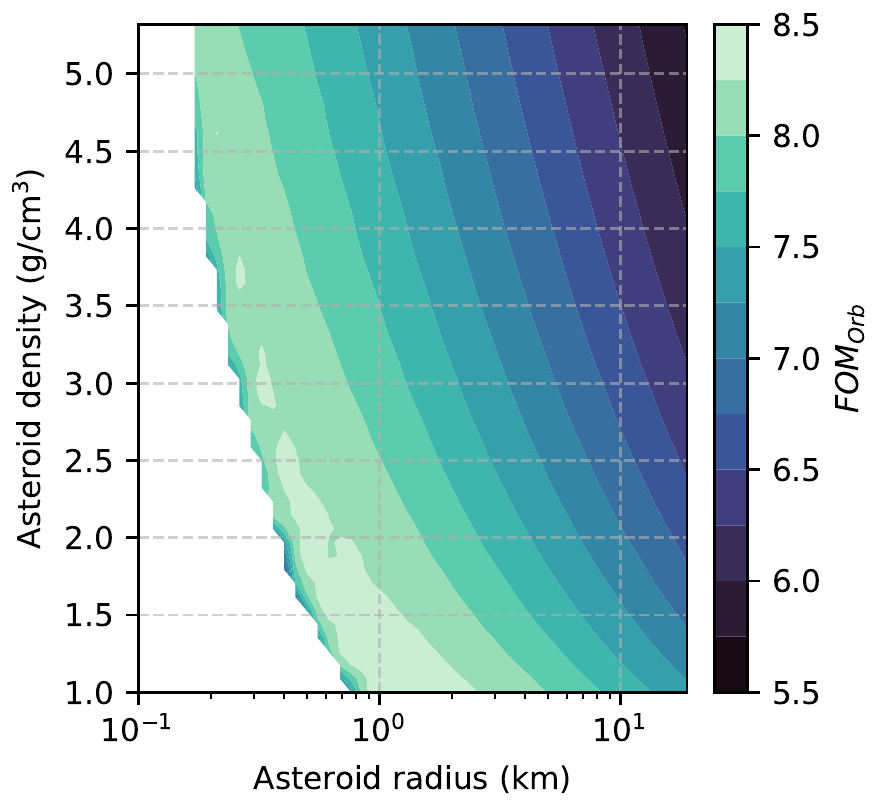}
		\caption{}
		\label{fig:fom_wcb_orbit}
	\end{subfigure}
	\hfill
	\begin{subfigure}[t]{0.45\textwidth}
		\centering
		\includegraphics[width=\textwidth]{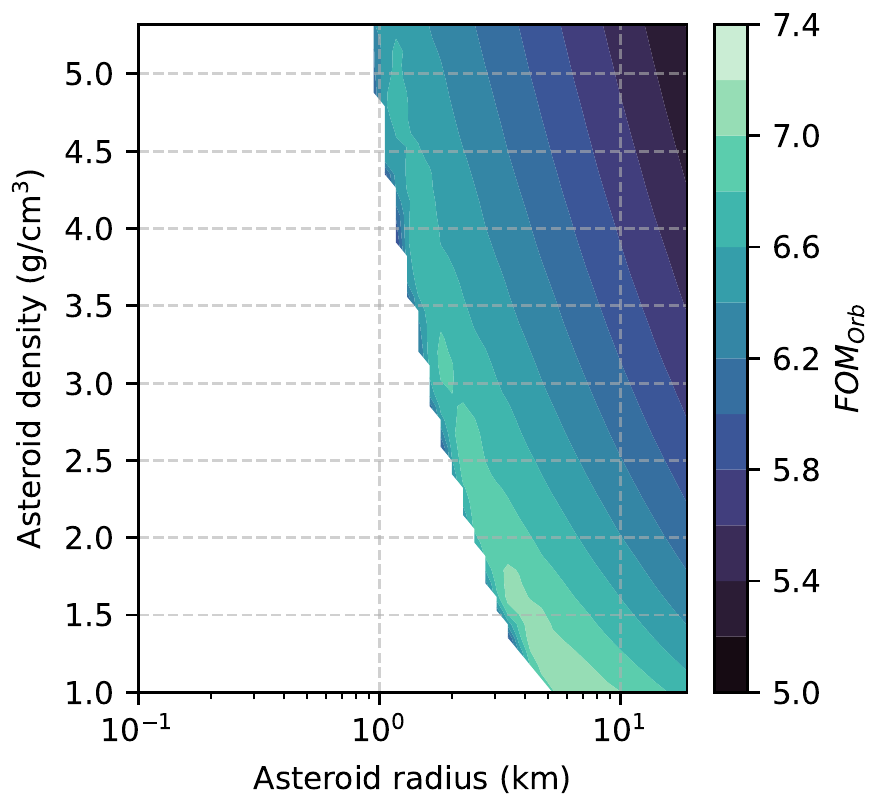}
		\caption{}
		\label{fig:fom_sfa_orbit}
	\end{subfigure}
	\caption{\label{fig:fom_plot_orbit}FOM as function of the target size and density for the orbiting collection strategy. (a) Sand-like material ($Y$ = 0 \si{\pascal}). (b) WCB ($Y$ = 1 \si{\kilo\pascal}). (c) SFA ($Y$ = 4 \si{\kilo\pascal})}
\end{figure}

In this section we present the analysis for the orbiting collection strategy. Similarly to \cref{subsec:l2_results}, we compare the FOM maps for the same three material cases (\cref{fig:fom_plot_orbit}). We can observe a similar behaviour between the materials, but with some differences. The first feature we can observe is the presence of an infeasibility region for the cases of WCB and SFA materials, while we do not observe this characteristic for the sand-like material. In this last case, all combinations of radius and density lead to particles availability for collection. This is a consequence of the relation between the minimum ejection speed and the strength of the material: the higher the strength the greater is the minimum ejection speed (\cref{eq:ejecta_speed}). Therefore, for smaller and less dense asteroids for which the escape velocity is small, the particles cannot stay trapped around the asteroid for a sufficient time to be collected. However, this behaviour also causes the second notable feature of the plots. In fact, it is possible to observe in \cref{fig:fom_wcb_orbit,fig:fom_sfa_orbit} that the highest FOM index is concentrated close to the border with the infeasible region. This is a consequence of the fact that the ejecta distribution has its peak close to the minimum velocity \cref{fig:ejecta_dist}. Therefore, if the required ejection speed is close to the minimum one, a higher number of particles will be available for collection.

\begin{figure}[htb!]
	\centering
	\includegraphics[width=3.4in]{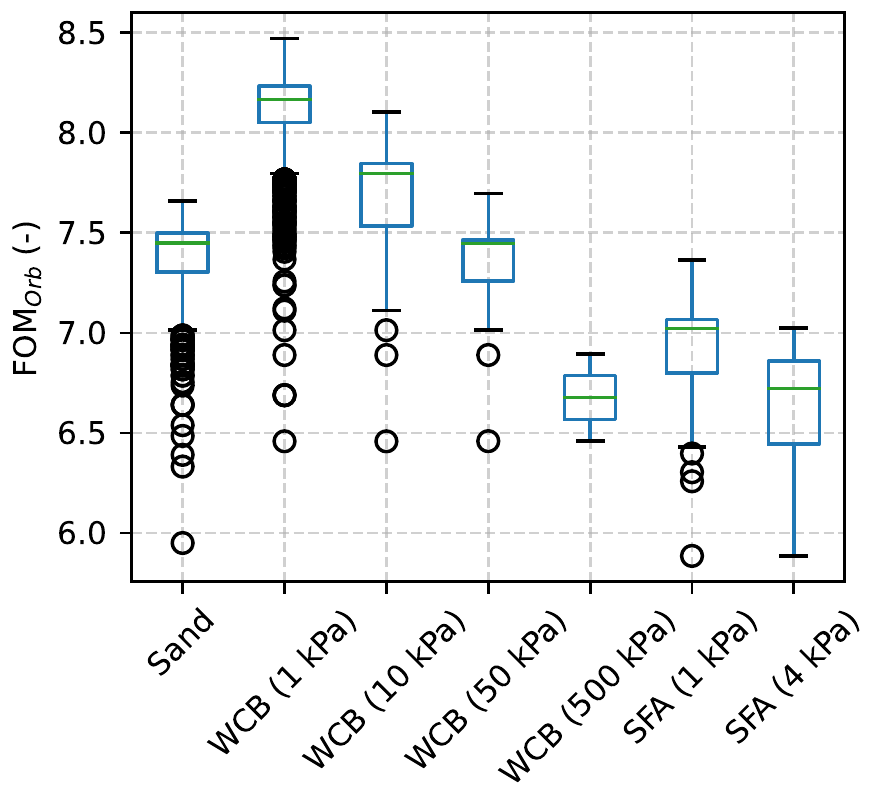}
	\caption{Boxplot of the $\mathrm{FOM}_{orb}$ as function of the material type and strength.}
	\label{fig:boxplot_fom_orbit}
\end{figure}

Differently from \cref{subsec:l2_results}, for the Orbiting case, the potential targets with the highest Figure of Merit correspond to solutions with smaller diameters and lower densities (the $L_2$ collection preferred high densities ones). This feature can be clearly observed in \cref{fig:fom_wcb_orbit,fig:fom_sfa_orbit}, with the maximum corresponding to the bottom left part of the feasible region, while \cref{fig:fom_sand_orbit} shows a maximum for low density asteroids with radii around 1 \si{\kilo\meter}. Given that a trend similar to the $L_2$ collection strategy is present here for what concerns the location of the infeasibility region, we can extend the discussion of \cref{subsec:l2_results} also to this case. Therefore, also for this collection strategy, we would prefer larger targets for a more robust target selection procedure; however, in this case, we would focus on low-density options instead of high-density ones.
\bigbreak
Specular to \cref{subsec:l2_results}, \cref{fig:boxplot_fom_orbit} shows the boxplot for the $\mathrm{FOM}_{orb}$ index, comparing the different materials and strengths we considered in the analysis. We can clearly observe a trend that matches the one of \cref{fig:boxplot_fom_l2}, with only minor differences. Particularly, a higher gap between the Figure of Merit of sand-like material and WCB, and a lower gap between the high-strength WCB (500 \si{\kilo\pascal}) and the remaining cases. As we are using two different methodologies to compute the FOM for the two collection strategies, we can use the results to compare different target solutions separately for the two methodologies, but we cannot the two methodologies among each other. The superiority of one collection methodology over the other will be mainly influenced by the operational constraints of the specific mission in exam.

\begin{figure}[htb!]
	\centering
	\includegraphics[width=3.4in]{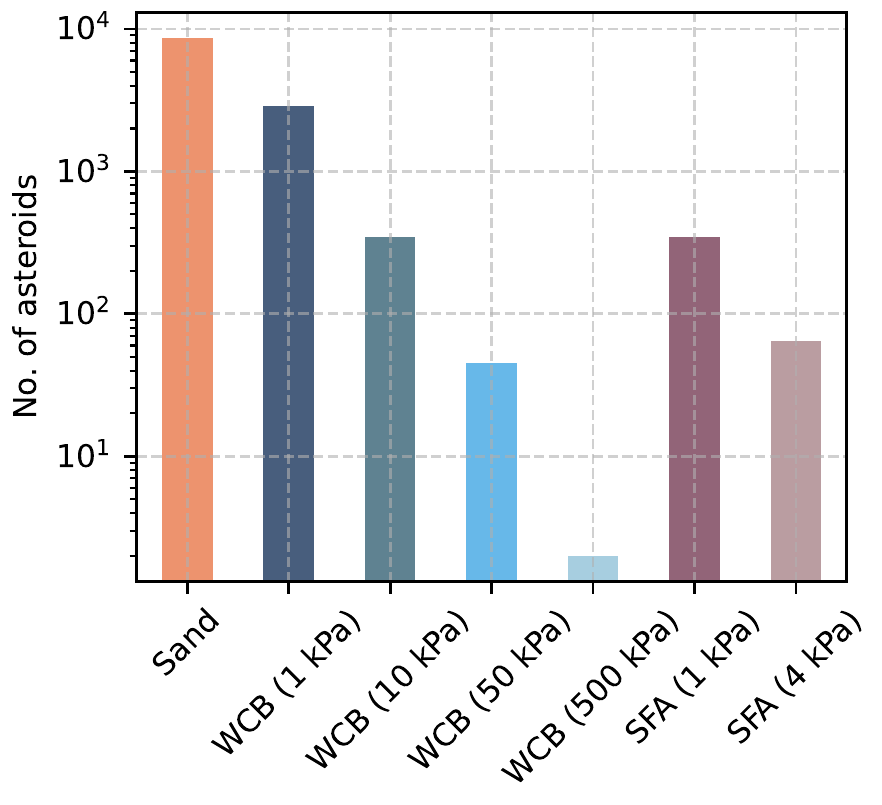}
	\caption{Number of identified targets for Orbiting collection as function of target material. Test particle: 1 \si{\milli\meter} size.}
	\label{fig:barplot_count_orbit}
\end{figure}

\cref{fig:barplot_count_orbit} shows also for the orbiting collection case the number of identified targets as a function of the target material and strength level. In general, the trend for the orbiting collection follows the trend of the $L_2$ collection of \cref{fig:barplot_count}, with a decreasing number of targets as the strength of the material increases. The same difference can also be seen between the different types of materials, with the low-strength SFA options giving results comparable to higher strengths WCB. Both these behaviours are driven by the impact physics and modelling and, specifically, by the resulting minimum ejection speed. Higher strengths lead to higher minimum ejection speeds, which eventually grow bigger than the escape velocity of the asteroid. Therefore, as the strength increases, a growing number of asteroids have escape speeds lower than the minimum ejection speed predicted by the ejecta model, which in turns lead to a lower number of possible targets.
Nonetheless, the number of identified targets for the orbiting collection scenario is greater than the one obtained for the $L_2$ options. Specifically, we observe a 120\% increase for sand-like materials, an average 20\% increase for WCB (with a peak of 33\% for the 1 \si{\kilo\pascal} case), and an average 22\% increase for SFA.

Equivalently to \cref{tab:rank_l2}, \cref{tab:rank_orb} shows a shortlist of the best identified targets, ranked based on the estimated $\Delta v$ required for the mission. Again, we have highlighted in bold the asteroids with the highest FOM index. We can observe that the table for the orbiting collection strategy closely resembles the ranking of the $L_2$ collection strategy, with only two notable differences: asteroids Wilson-Harrington and Pygmalion. The asteroid Wilson-Harrington has also the highest Figure of Merit, indicating the highest potential for collection for the orbiting strategy.

A final table is presented in the following (\cref{tab:rank_fom_orb}), showing a shortlist of asteroid ranked based on the $\mathrm{FOM}_{orb}$ index, instead of the $\Delta v$. In this case, we observe that the ranking considerably differs from the  $\Delta v$-based one of \cref{tab:rank_orb}, introducing several new potential targets. However, in contrast with the previous two tables, one-third of the identified targets have large $\Delta v$ requirements (more than 10 \si{\kilo\meter\per\second}) and are therefore very expensive to reach.

\begin{table}[htb!]
	\centering
	\caption{\label{tab:rank_orb} Target asteroid ranking for the orbiting collection strategy. Ranking from the lowest $\Delta v$.}
	\begin{tabular}{lccccccc}
		\hline
		Name & Sand & WCB & WCB & WCB & SFA & SFA & $\Delta v$ \\
		& & 1 kPa & 10 kPa & 50 kPa  & 1 kPa & 4 kPa & (km/s) \\
		\hline
		\textbf{Anteros} & \textbf{7.25} &        \textbf{7.93} &         \textbf{7.86} &          - &        \textbf{7.08} &         - &      \textbf{5.11} \\
		Eros & 6.33 &        6.89 &         6.89 &         6.89 &        6.26 &        6.26 &      5.58 \\
		\textbf{Zephyr} & \textbf{7.29} &        \textbf{7.98} &         \textbf{7.65} &          - &        \textbf{6.87} &         - &      \textbf{5.86} \\
		Geographos & 7.21 &        7.89 &         7.93 &          - &        7.15 &         - &      6.08 \\
		Ivar & 6.64 &        7.24 &         7.24 &         7.24 &        6.56 &        6.56 &      6.15 \\
		McAuliffe & 7.17 &        7.84 &         7.86 &          - &        7.09 &         - &      6.18 \\
		Seleucus & 6.96 &        7.63 &         7.62 &          - &        6.86 &         - &      6.20 \\
		Oze & 6.74 &        7.37 &         7.37 &         7.36 &        6.65 &        6.63 &      6.33 \\
		Toro & 7.08 &        7.74 &         7.73 &          - &        6.98 &         - &      6.43 \\
		Toutatis & 6.89 &        7.52 &         7.52 &         7.51 &        6.80 &        6.85 &      6.50 \\
		Melissabrucker & 6.84 &        7.47 &         7.47 &         7.46 &        6.75 &        6.74 &      6.60 \\
		\begin{tabular}{@{}l@{}} \textbf{Wilson-} \\ \textbf{Harrington} \end{tabular} & \textbf{7.53} &        \textbf{8.20} &         \textbf{7.81} &          - &        \textbf{7.07} &         - &      \textbf{6.61} \\
		Beltrovata & 7.25 &        7.93 &         7.86 &          - &        7.08 &         - &      6.66 \\
		\textbf{Cacus} & \textbf{7.29} &        \textbf{7.98} &         \textbf{7.65} &          - &        \textbf{6.87} &         - &      \textbf{6.93} \\
		Pygmalion & 7.06 &        7.71 &         7.71 &          - &        6.97 &        5.96 &      6.96 \\
		Alinda & 6.99 &        7.63 &         7.63 &         7.10 &        6.90 &        6.76 &      6.96 \\
		\hline
	\end{tabular}
\end{table}

\begin{table}[htb!]
	\centering
	\caption{\label{tab:rank_fom_orb} Target asteroid ranking for the orbiting collection strategy. Ranking from the highest $\mathrm{FOM}_{orb}$.}
	\begin{tabular}{lrrrrrrr}
		\hline
		Name & Sand & WCB & WCB & WCB & SFA & SFA & $\Delta v$ \\
		& & 1 kPa & 10 kPa & 50 kPa  & 1 kPa & 4 kPa & (km / s)\\
		\hline
      	Davidharvey & 7.53 &        8.20 &         7.81 &          - &        7.07 &         - &      7.18 \\
		\begin{tabular}{@{}l@{}} \textbf{Wilson-} \\ \textbf{Harrington} \end{tabular} & 7.53 &        8.20 &         7.81 &          - &        7.07 &         - &      \textbf{6.61} \\
		Heracles & 7.50 &        8.16 &         8.06 &          - &        7.31 &         - &      9.52 \\
		Betulia & 7.44 &        8.08 &         8.10 &          - &        7.36 &         - &     14.75 \\
		Midas & 7.34 &        8.01 &         7.97 &          - &        7.21 &         - &     12.65 \\
		Izhdubar & 7.29 &        7.98 &         7.65 &          - &        6.87 &         - &     15.10 \\
		Cacus & 7.29 &        7.98 &         7.65 &          - &        6.87 &         - &      6.93 \\
		\textbf{Zephyr} & 7.29 &        7.98 &         7.65 &          - &        6.87 &         - &      \textbf{5.86} \\
		Cruithne & 7.29 &        7.98 &         7.65 &          - &        6.87 &         - &     13.82 \\
		Ondaatje & 7.29 &        7.98 &         7.65 &          - &        6.87 &         - &      7.65 \\
		\textbf{Anteros} & 7.25 &        7.93 &         7.86 &          - &        7.08 &         - &      \textbf{5.11} \\
		\hline
	\end{tabular}
\end{table}

\subsection{Preliminary risk analysis from particle impacts}  \label{subsec:risk_analysis}
\begin{figure}[htb!]
	\centering
	\includegraphics[width=2.6in]{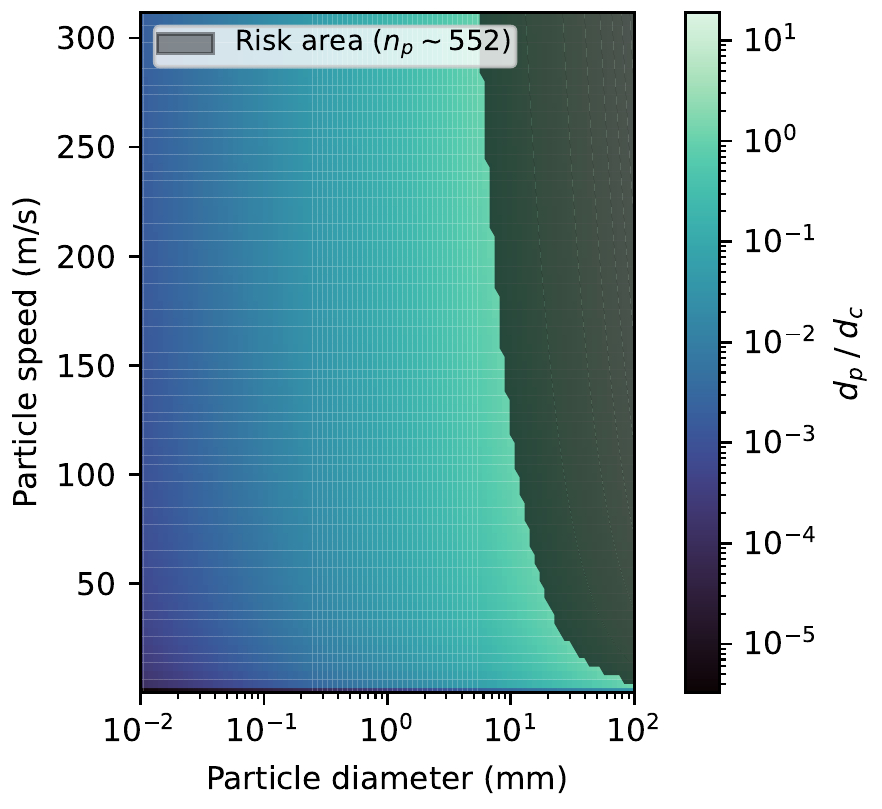}
	\caption{Ratio between particle diameter and critical diameter for sand-like materials. Black shaded area: risk region.}
	\label{fig:dc_dp_sand}
\end{figure}

\begin{figure}[htb!]
	\centering
	\includegraphics[width=2.6in]{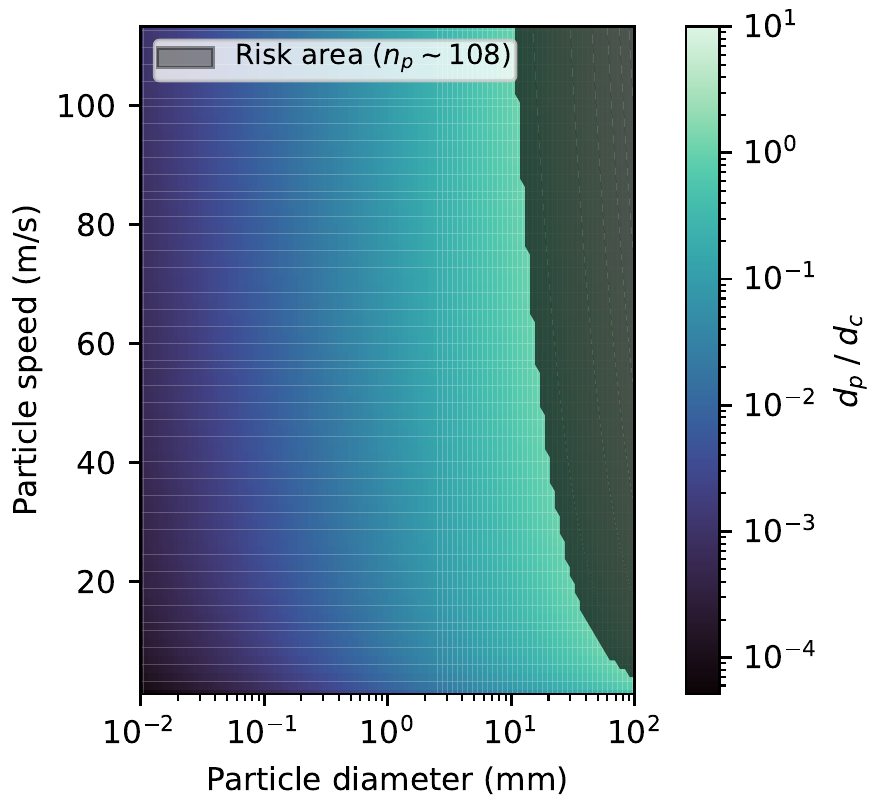}
	\caption{Ratio between particle diameter and critical diameter for WCB material. Black shaded area: risk region.}
	\label{fig:dc_dp_wcb}
\end{figure}

To assess the feasibility of the mission it is also necessary to verify that the impact of the particles with the spacecraft is not dangerous to compromise the mission. For this assessment, we model the impact of the orbiting particles with the spacecraft using Ballistic Limit Equations (BLEs) \citep{Shannon2009}, which have been extensively used to evaluate the risk posed by space debris on objects orbiting the Earth \citep{TC2020VulnerableZones}. For this study, a conservative approach has been followed. We consider a single-wall BLE for ballistic impacts (below the hypervelocity limit). We use the BLE to derive the value of the critical particle diameter that is the minimum particle diameter causing damage on a single-wall structure. The corresponding equation is the following:

\begin{equation}
	\label{eq:ble}
	d_c = \left[ \frac{\frac{1}{K_{3S}} \cdot t_w^{0.5} \cdot \left( \frac{\sigma_y}{40} \right)^{1/2}}{0.6 \cdot \left( \cos \theta \right)^\delta \rho_p^{1/2} u_p^{2/3}} \right]^{18/19}
\end{equation}

where $\theta$ is the impact angle (between the particle velocity and the normal to the considered surface), $d_c$ is the critical diameter required to damage the single wall plate of thickness $t_w$ if it is hit by a particle of density $\rho_p$ and velocity $u_p$. For the case in exam, we consider an aluminium plate with tensile strength $\sigma_y$ = 276 \si{\mega\pascal} (to be converted in ksi for use in \cref{eq:ble}) and a normal impact ($\theta$ = 0 deg). The parameters $K_{3S}$ = 1.4 and $\delta$ = 4/3 are associated to impacts on aluminium plates \citep{Shannon2009}. Combining the ejecta distribution of \cref{eq:ejecta_dist} with the information of the BLEs, we can estimate which ejected particles can be dangerous for the spacecraft that is all particles with a diameter larger than the critical diameter as predicted by \cref{eq:ble}). \cref{fig:dc_dp_sand,fig:dc_dp_wcb} show the variation of the ratio between the particle diameter and the critical diameter for the ejecta distributions relative to sand-like material and WCB, respectively.

In both cases we can observe a black shaded region concentrated in the upper right portion of the plot that identifies potentially dangerous particles. In this case the particle diameter is greater than the critical one. As expected, the larger the particle the more likely is to cause damage even at smaller velocities. We can also observe that the sand-like material \cref{fig:dc_dp_sand} has a larger risk area since the ejection velocities are higher than the WCB case. In addition, by combining the shaded area of \cref{fig:dc_dp_sand,fig:dc_dp_wcb} with the particle density distribution (equivalent to \cref{fig:ejecta_dist})), we can estimate the number of dangerous particles for both cases. We can observe that a sand-like material generates five time the dangerous particles of WCB, with about 550 risky fragments. In addition, we observe that at low velocities, no particle diameter is dangerous for the spacecraft and as the ejection speed increases, particle diameters down to 10 \si{\milli\meter} can cause damages.

\section{Conclusions and Discussion}
\label{sec:conclusions}

In this work, we focused on understanding the contribution of the asteroid characteristics to the selection of suitable targets for in-orbit sample collection missions. The outcome of the analyses stems from the combination of the particle dynamics and the ejecta modelling. Throughout the work, we have considered two collection scenarios: a first approach that seeks to collect those particles in the anti-solar direction, exploiting the features of the L$_2$ Lagrangian point, and a second approach that instead seeks to gather the particles residing in the neighbourhood of the asteroid for a sufficient time. For both collection options, a Figure of Merit has been defined for a preliminary assessment of the potential collection capabilities, and has been evaluated as function of the target and material properties. Indeed, the FOM is directly related to an estimate of the potentially collectable particles. For both scenarios, feasibility regions have been identified considering relevant characteristics of the target (i.e., the asteroid radius and density). Additionally, uncertain parameters such as the material type and strength of the asteroid have been taken into account. From the results of \cref{subsec:orbit_results}, the orbiting collection is favoured by sand-like materials as no infeasible regions can be identified. Infeasibility regions are instead present in all the remaining analysed cases, for both collection options. These regions grow as the strength of the analysed materials grow. For comparable strength levels, they appear more predominant for the SFA material rather than WCB.
Beside the collection feasibility, the FOM index gives also an approximate estimate of the potentially available number of particles. This is achieved by combining the understanding of the particles' dynamics with the development of a distribution-based ejecta model. The maps of \cref{fig:fom_plot_l2,fig:fom_plot_orbit} show that within the feasibility region, both collection options can be more efficient for smaller asteroids; however, the L$_2$ option favours denser targets, while the orbiting options less dense ones.
The generated maps and the introduced FOMs have been applied to the search of potential targets by scanning the Near-Earth Asteroid database. By placing inside the derived maps the NEAs, based on their radius and estimated density, we can assign them a value of the FOM to evaluate their potential for in-orbit collection.
were used to scan the Near-Earth Asteroid database to identify targets and possibly rank them. For both the collections scenarios thousands and hundreds of potential target were identified, when considering low-strength materials. However, as the strength increases, fewer targets remain available: around 40 targets for the 50 \si{\kilo\pascal} WCB and only a couple for the 500 \si{\kilo\pascal} case. Combining the maps of \cref{fig:fom_plot_l2,fig:fom_plot_orbit} and estimating the $\Delta v$ required to rendezvous with the asteroids, we also ranked the potential target asteroids as function of both the minimum required $\Delta v$ and the maximum FOM. Performing the target selection analysis for different material types and strengths has also been fundamental to improve the robustness of the target selection procedure. In fact, given the uncertainties in the knowledge of the properties of the asteroid soil, a more robust selection can be achieved by looking at those targets that show potential for collection for more combinations of material types and strengths.

With the maps of \cref{sec:results}, we wanted to prioritise the understanding of the effects of the asteroid characteristics, which also have a strong influence on the ejecta generation process. However, it is important to remark that the methodology used is simplified in nature and is influenced by a set of assumptions. Therefore, the maps represent a useful tool to perform a preliminary selection of the targets that is to obtain a shortlist of candidates for which more refined analysis should be carried out. For example, the assumption of a constant semi-major axis can have an influence on the target evaluation as the asteroids in the NEA database will deviate from this value. In this work, we have not considered this parameter in order to limit the number of simulations, while still retaining an understanding of the influence of other fundamental parameters. Additional maps could be generated for different semi-major axes to have a more refined target evaluation. Once the target of the mission is selected, dedicated simulations should be performed, considering the specificity of the dynamical environment of the selected asteroid. A first analysis on the candidate impact locations should be performed, based on the type of collection strategy and the operational constraints. Once the impact site is selected, detailed simulations of the ejecta fate should be performed to identify the better collection locations, for example, evaluating the flux of particles the spacecraft can expect given its location and orientation. Given the dependency of the analyses on several parameters, multiple runs should be performed for a more statistically-based collection analysis.

In this work, the operational constraints and requirements for the proposed mission concept have not been tackled in detail. \cref{subsec:risk_analysis} provides a preliminary analysis and discussion on the potential risk for the mission. However, other operational constraints will need to be considered for the future development of the mission concept. As an example, the L$_2$ collection methodology requires the spacecraft to be positioned in the anti-solar direction, which can introduce challenges in maintaining the spacecraft in such position and can expose the spacecraft to long eclipse periods. Future work will be dedicated to more detailed analyses of both collection strategies, analysing their strength, weaknesses and operational constraints for a better understanding of the feasibility of the proposed mission concept. However, given the preliminary nature of the analysis, such operational constraints have not been accounted for and will be the subject of future studies.
In fact, future works will farther the analysis on the proposed in-orbit particle collection mission concept, detailing the operational phases of the identified collection strategies and providing a more detail analysis of their particle collection efficacy. In addition, the preliminary design of the spacecraft and of the collection device will be performed.

Finally, a preliminary assessment of the risk the ejected particles pose to the spacecraft showed that few particles (in the order of hundreds) can be dangerous. This is quite a small value, compared to the total number of fragments generated by an impact. Moreover, the performed analysis is conservative as a 1 mm single-wall configuration has been analysed. More common options, such as honeycomb sandwich panels can offer a greater protection and further reduce the risk for the spacecraft.

\section*{Acknowledgements}
This project has received funding from the European Union’s Horizon 2020 research and innovation programme under the Marie Sklodowska-Curie grant agreement No 896404 - CRADLE.

\bibliographystyle{elsarticle-num}
\bibliography{references}

\begin{thebibliography}{10}
\expandafter\ifx\csname url\endcsname\relax
  \def\url#1{\texttt{#1}}\fi
\expandafter\ifx\csname urlprefix\endcsname\relax\def\urlprefix{URL }\fi
\expandafter\ifx\csname href\endcsname\relax
  \def\href#1#2{#2} \def\path#1{#1}\fi

\bibitem{hein2020techno}
A.~M. Hein, R.~Matheson, D.~Fries, {A techno-economic analysis of asteroid
  mining}, Acta Astronautica 168 (2020) 104--115.
\newblock \href {http://dx.doi.org/10.1016/j.actaastro.2019.05.009}
  {\path{doi:10.1016/j.actaastro.2019.05.009}}.

\bibitem{xie2021target}
R.~Xie, N.~J. Bennett, A.~G. Dempster, {Target evaluation for near earth
  asteroid long-term mining missions}, Acta Astronautica 181 (2021) 249--270.
\newblock \href {http://dx.doi.org/10.1016/j.actaastro.2021.01.011}
  {\path{doi:10.1016/j.actaastro.2021.01.011}}.

\bibitem{kawaguchi2008hayabusa}
J.~Kawaguchi, A.~Fujiwara, T.~Uesugi, {Hayabusa—Its technology and science
  accomplishment summary and Hayabusa-2}, Acta Astronautica 62~(10-11) (2008)
  639--647.
\newblock \href {http://dx.doi.org/10.1016/j.actaastro.2008.01.028}
  {\path{doi:10.1016/j.actaastro.2008.01.028}}.

\bibitem{tsuda2013system}
Y.~Tsuda, M.~Yoshikawa, M.~Abe, H.~Minamino, S.~Nakazawa, {System design of the
  Hayabusa 2—Asteroid sample return mission to 1999 JU3}, Acta Astronautica
  91 (2013) 356--362.
\newblock \href {http://dx.doi.org/10.1016/j.actaastro.2013.06.028}
  {\path{doi:10.1016/j.actaastro.2013.06.028}}.

\bibitem{tsuda2019hayabusa2}
Y.~Tsuda, M.~Yoshikawa, T.~Saiki, S.~Nakazawa, S.-i. Watanabe,
  {Hayabusa2--Sample return and kinetic impact mission to near-earth asteroid
  Ryugu}, Acta Astronautica 156 (2019) 387--393.
\newblock \href {http://dx.doi.org/10.1016/j.actaastro.2018.01.030}
  {\path{doi:10.1016/j.actaastro.2018.01.030}}.

\bibitem{tsuda2020hayabusa2}
Y.~Tsuda, T.~Saiki, F.~Terui, S.~Nakazawa, M.~Yoshikawa, S.-i. Watanabe, H.~P.
  Team, {Hayabusa2 mission status: Landing, roving and cratering on asteroid
  Ryugu}, Acta Astronautica 171 (2020) 42--54.
\newblock \href {http://dx.doi.org/10.1016/j.actaastro.2020.02.035}
  {\path{doi:10.1016/j.actaastro.2020.02.035}}.

\bibitem{witte2016rosetta}
L.~Witte, R.~Roll, J.~Biele, S.~Ulamec, E.~Jurado, {Rosetta lander
  Philae-Landing performance and touchdown safety assessment}, Acta
  Astronautica 125 (2016) 149--160.
\newblock \href {http://dx.doi.org/10.1016/j.actaastro.2016.02.001}
  {\path{doi:10.1016/j.actaastro.2016.02.001}}.

\bibitem{ulamec2016rosetta}
S.~Ulamec, C.~Fantinati, M.~Maibaum, K.~Geurts, J.~Biele, S.~Jansen,
  O.~K{\"u}chemann, B.~Cozzoni, F.~Finke, V.~Lommatsch, et~al., {Rosetta
  lander--landing and operations on comet 67P/Churyumov--Gerasimenko}, Acta
  Astronautica 125 (2016) 80--91.
\newblock \href {http://dx.doi.org/10.1016/j.actaastro.2015.11.029}
  {\path{doi:10.1016/j.actaastro.2015.11.029}}.

\bibitem{blume2003deep}
W.~H. Blume, {Deep Impact: mission design approach for a new Discovery
  mission}, Acta Astronautica 52~(2-6) (2003) 105--110.
\newblock \href {http://dx.doi.org/10.1016/S0094-5765(02)00144-3}
  {\path{doi:10.1016/S0094-5765(02)00144-3}}.

\bibitem{lauretta2017osiris}
D.~Lauretta, S.~Balram-Knutson, E.~Beshore, W.~Boynton, C.~D. d’Aubigny,
  D.~DellaGiustina, H.~Enos, D.~Golish, C.~Hergenrother, E.~Howell, et~al.,
  {OSIRIS-REx: sample return from asteroid (101955) Bennu}, Space Science
  Reviews 212~(1) (2017) 925--984.
\newblock \href {http://dx.doi.org/10.1007/s11214-017-0405-1}
  {\path{doi:10.1007/s11214-017-0405-1}}.

\bibitem{dellagiustina2022osiris}
D.~DellaGiustina, D.~Golish, S.~Guzewich, M.~Moreau, M.~Nolan, A.~Polit,
  A.~Simon, Osiris-apex: A proposed osiris-rex extended mission to apophis, LPI
  Contributions 2681 (2022) 2011.

\bibitem{2019adamsDART}
E.~Adams, D.~O'Shaughnessy, M.~Reinhart, J.~John, E.~Congdon, D.~Gallagher,
  E.~Abel, J.~Atchison, Z.~Fletcher, M.~Chen, C.~Heistand, P.~Huang, E.~Smith,
  D.~Sibol, D.~Bekker, D.~Carrelli, Double asteroid redirection test: The earth
  strikes back, in: 2019 IEEE Aerospace Conference, 2019, pp. 1--11.
\newblock \href {http://dx.doi.org/10.1109/AERO.2019.8742007}
  {\path{doi:10.1109/AERO.2019.8742007}}.

\bibitem{latino2019iac}
A.~Latino, S.~Soldini, C.~Colombo, Y.~Tsuda, et~al., {Ejecta orbital and
  bouncing dynamics around asteroid Ryugu}, in: 70th International
  Astronautical Congress (IAC 2019), 2019, pp. 1--19.

\bibitem{latino2019thesis}
A.~Latino, Ejecta orbital and bouncing dynamics around asteroid ryugu, Master's
  thesis (2019).

\bibitem{trisolini2021ejecta}
M.~Trisolini, C.~Colombo, Y.~Tsuda, et~al., {Ejecta dynamics around asteroids
  in view of in-orbit particle collection missions}, in: 72nd International
  Astronautical Congress (IAC 2021), 2021, pp. 1--10.

\bibitem{trisolini2022scitech}
M.~Trisolini, C.~Colombo, Y.~Tsuda, {Ejecta models for particles generated by
  small kinetic impactors onto asteroid surfaces}, in: AIAA SCITECH 2022 Forum,
  2022, p. 2383.
\newblock \href {http://dx.doi.org/10.2514/6.2022-2383}
  {\path{doi:10.2514/6.2022-2383}}.

\bibitem{ulamec2014landing}
S.~Ulamec, J.~Biele, P.-W. Bousquet, P.~Gaudon, K.~Geurts, T.-M. Ho, C.~Krause,
  C.~Lange, R.~Willnecker, L.~Witte, et~al., {Landing on small bodies: From the
  Rosetta Lander to MASCOT and beyond}, Acta Astronautica 93 (2014) 460--466.
\newblock \href {http://dx.doi.org/10.1016/j.actaastro.2013.02.007}
  {\path{doi:10.1016/j.actaastro.2013.02.007}}.

\bibitem{sachse2015correlation}
M.~Sachse, J.~Schmidt, S.~Kempf, F.~Spahn, {Correlation between speed and size
  for ejecta from hypervelocity impacts}, Journal of Geophysical Research:
  Planets 120~(11) (2015) 1847--1858.
\newblock \href {http://dx.doi.org/10.1002/2015JE004844}
  {\path{doi:10.1002/2015JE004844}}.

\bibitem{housen2011ejecta}
K.~R. Housen, K.~A. Holsapple, {Ejecta from impact craters}, Icarus 211~(1)
  (2011) 856--875.
\newblock \href {http://dx.doi.org/10.1016/j.icarus.2010.09.017}
  {\path{doi:10.1016/j.icarus.2010.09.017}}.

\bibitem{holsapple2007crater}
K.~A. Holsapple, K.~R. Housen, {A crater and its ejecta: An interpretation of
  Deep Impact}, Icarus 191~(2) (2007) 586--597.
\newblock \href {http://dx.doi.org/10.1016/j.icarus.2006.08.035}
  {\path{doi:10.1016/j.icarus.2006.08.035}}.

\bibitem{scheeres2002fate}
D.~Scheeres, D.~Durda, P.~Geissler, The fate of asteroid ejecta, Asteroids III
  1 (2002) 527--544.

\bibitem{scheeres2000temporary}
D.~Scheeres, F.~Marzari, Temporary orbital capture of ejecta from comets and
  asteroids: Application to the deep impact experiment, Astronomy and
  Astrophysics 356 (2000) 747--756.

\bibitem{richter1995stability}
K.~Richter, H.~Keller, On the stability of dust particle orbits around cometary
  nuclei, Icarus 114~(2) (1995) 355--371.

\bibitem{yu2017ejecta}
Y.~Yu, P.~Michel, S.~R. Schwartz, S.~P. Naidu, L.~A. Benner, {Ejecta cloud from
  the AIDA space project kinetic impact on the secondary of a binary asteroid:
  I. mechanical environment and dynamical model}, Icarus 282 (2017) 313--325.
\newblock \href {http://dx.doi.org/10.1016/j.icarus.2016.09.008}
  {\path{doi:10.1016/j.icarus.2016.09.008}}.

\bibitem{yu2018ejecta}
Y.~Yu, P.~Michel, {Ejecta cloud from the AIDA space project kinetic impact on
  the secondary of a binary asteroid: II. Fates and evolutionary dependencies},
  Icarus 312 (2018) 128--144.
\newblock \href {http://dx.doi.org/10.1016/j.icarus.2018.04.017}
  {\path{doi:10.1016/j.icarus.2018.04.017}}.

\bibitem{rossi2022dynamical}
A.~Rossi, F.~Marzari, J.~R. Brucato, V.~Della~Corte, E.~Dotto, S.~Ieva, S.~L.
  Ivanovski, A.~Lucchetti, E.~M. Epifani, M.~Pajola, et~al., Dynamical
  evolution of ejecta from the dart impact on dimorphos, The Planetary Science
  Journal 3~(5) (2022) 118.

\bibitem{soldini2017assessing}
S.~Soldini, Y.~Tsuda, {Assessing the hazard posed by Ryugu ejecta dynamics on
  Hayabusa2 spacecraft}, in: 26th International Symposium of Space Flight
  Dynamics, 2017, pp. 1--11.

\bibitem{pinto2020}
D.~V. Pinto, S.~Soldini, Y.~Tsuda, J.~Heiligers, {Temporary capture of asteroid
  ejecta into periodic orbits: Application to jaxa's hayabusa2 impact event},
  AIAA Scitech 2020 Forum 1 PartF~(January) (2020) 1--21.
\newblock \href {http://dx.doi.org/10.2514/6.2020-0221}
  {\path{doi:10.2514/6.2020-0221}}.

\bibitem{hasegawa2018physical}
S.~Hasegawa, D.~Kuroda, K.~Kitazato, T.~Kasuga, T.~Sekiguchi, N.~Takato,
  K.~Aoki, A.~Arai, Y.-J. Choi, T.~Fuse, et~al., {Physical properties of
  near-Earth asteroids with a low delta-v: Survey of target candidates for the
  Hayabusa2 mission}, Publications of the Astronomical Society of Japan 70~(6)
  (2018) 114.
\newblock \href {http://dx.doi.org/10.1093/pasj/psy119}
  {\path{doi:10.1093/pasj/psy119}}.

\bibitem{scheeres2016orbital}
D.~J. Scheeres, Orbital motion in strongly perturbed environments: applications
  to asteroid, comet and planetary satellite orbiters, Springer, 2016.

\bibitem{vallado2001fundamentals}
D.~A. Vallado, Fundamentals of astrodynamics and applications, Vol.~12,
  Springer Science \& Business Media, 2001.

\bibitem{nasa_sbdb}
NASA, Nasa small-body database,
  \url{https://ssd.jpl.nasa.gov/tools/sbdb_query.html}, accessed: December 2021
  (2021).

\bibitem{szebehely2012theory}
V.~Szebehely, Theory of orbit: The restricted problem of three Bodies,
  Elsevier, 2012.

\bibitem{arakawa2020artificial}
M.~Arakawa, T.~Saiki, K.~Wada, K.~Ogawa, T.~Kadono, K.~Shirai, H.~Sawada,
  K.~Ishibashi, R.~Honda, N.~Sakatani, et~al., {An artificial impact on the
  asteroid (162173) Ryugu formed a crater in the gravity-dominated regime},
  Science 368~(6486) (2020) 67--71.
\newblock \href {http://dx.doi.org/10.1126/science.aaz1701}
  {\path{doi:10.1126/science.aaz1701}}.

\bibitem{saiki2013small}
T.~Saiki, H.~Sawada, C.~Okamoto, H.~Yano, Y.~Takagi, Y.~Akahoshi, M.~Yoshikawa,
  {Small carry-on impactor of Hayabusa2 mission}, Acta Astronautica 84 (2013)
  227--236.
\newblock \href {http://dx.doi.org/10.1016/j.actaastro.2012.11.010}
  {\path{doi:10.1016/j.actaastro.2012.11.010}}.

\bibitem{richardson2007ballistics}
J.~E. Richardson, H.~J. Melosh, C.~M. Lisse, B.~Carcich, {A ballistics analysis
  of the Deep Impact ejecta plume: Determining Comet Tempel 1's gravity, mass,
  and density}, Icarus 191~(2) (2007) 176--209.
\newblock \href {http://dx.doi.org/10.1016/j.icarus.2007.08.033}
  {\path{doi:10.1016/j.icarus.2007.08.033}}.

\bibitem{holsapple2012momentum}
K.~A. Holsapple, K.~R. Housen, {Momentum transfer in asteroid impacts. I.
  Theory and scaling}, Icarus 221~(2) (2012) 875--887.
\newblock \href {http://dx.doi.org/10.1016/j.icarus.2012.09.022}
  {\path{doi:10.1016/j.icarus.2012.09.022}}.

\bibitem{carry2012density}
B.~Carry, Density of asteroids, Planetary and Space Science 73~(1) (2012)
  98--118.
\newblock \href {http://dx.doi.org/10.1016/j.pss.2012.03.009}
  {\path{doi:10.1016/j.pss.2012.03.009}}.

\bibitem{trisolini2021jaxaws}
M.~Trisolini, C.~Colombo, Y.~Tsuda, {Ejecta analysis for an asteroid impact
  event in the perturbed circular restricted three body problem}, in: 31st JAXA
  Workshop on Astrodynamics and Flight Mechanics, 2021.

\bibitem{zanotti2019hypervelocity}
G.~Zanotti, Hypervelocity impacts on planetary bodies: Modelling craters
  formation and ejecta plume evolution, Master's thesis, Politecnico di Milano
  (2019).

\bibitem{matsumoto2011numerical}
J.~Matsumoto, T.~Saiki, Y.~Tsuda, J.~Kawaguchi, Numerical analysis of particle
  distribution with collisions around an asteroid, in: title The 21th Workshop
  on JAXA Astrodynamics and Flight Mechanics, 2011, p. 327.

\bibitem{shoemaker1978earth}
E.~Shoemaker, E.~Helin, {Earth-approaching asteroids as targets for
  exploration}, Asteroids: An exploration assessment (1978) 245--256.

\bibitem{Shannon2009}
R.~Shannon, E.~Christiansen, {Micrometeoroid and Orbital Debris (MMOD) Shield
  Ballistic Limit Analysis Program} (2009).

\bibitem{TC2020VulnerableZones}
M.~Trisolini, H.~G. Lewis, C.~Colombo, {Predicting the vulnerability of
  spacecraft components: Modelling debris impact effects through
  vulnerable-zones}, Advances in Space Research 65~(11) (2020) 2692--2710.
\newblock \href {http://dx.doi.org/10.1016/j.asr.2020.03.010}
  {\path{doi:10.1016/j.asr.2020.03.010}}.

\end{thebibliography}

\end{document}